\documentclass[preprint,review,12pt]{elsarticle}

\usepackage{amssymb}
\usepackage{amsmath}
\usepackage{graphicx}
\usepackage{longtable}
\usepackage[colorlinks=true]{hyperref}
\usepackage{cleveref}
\usepackage{natbib}
\usepackage{tikz}
\usepackage[normalem]{ulem}
\usetikzlibrary{fit,calc,patterns}
\usepackage{subcaption}
\usepackage{svg}
\usepackage[ruled,vlined]{algorithm2e}
\usepackage[dvipsnames]{xcolor}
\newcommand{\ts}[1]{\textsuperscript{#1}}
\newcommand{\myeqref}[1]{{\renewcommand{\equationautorefname}{Eq.}\autoref{#1}}}

\colorlet{mypink}{red!40}
\colorlet{myblue}{cyan!60}
\newcommand{\tc}[1]{\textcolor{blue}{#1}}

\renewcommand{\tc}[1]{#1}
\renewcommand{\sout}[1]{}

\usepackage{lineno}
\usepackage{import}
\usepackage{tabularx}
\journal{Chemical Engineering Journal}

\begin{document}

\begin{frontmatter}

\title{\tc{Multiphase modeling of anisotropic biomass particle pyrolysis accounting for particle deformation and coupled gas-phase dynamics}}

\author{Riccardo Caraccio}
\author{Edoardo Cipriano\fnref{fn1}} 
\author{Alessio Frassoldati} 
\author{Tiziano Faravelli\corref{cor1}}
\ead{tiziano.faravelli@polimi.it}
\fntext[fn1]{\textit{Present address}: Institut Jean Le Rond d’Alembert UMR 7190, Sorbonne Université and CNRS, Paris 75005, France}

\cortext[cor1]{Corresponding Author}
\affiliation{organization={CRECK Modeling Lab, Department of Chemistry, Materials and Chemical Engineering "G. Natta", Politecnico di Milano},
                     addressline={Piazza Leonardo Da Vinci 32},
                     city={Milano},
                     postcode={20133},
                     country={Italy}}

\begin{abstract}
Numerical models of biomass particle pyrolysis focus on either the solid particle evolution or on the surrounding gas-phase dynamics, neglecting the coupled interactions between the two. This work addresses this limitation by proposing a single-grid model that fully resolves both phases without relying on sub-grid-scale correlations. The model adopts an Eulerian representation of the two-phase system, using a Volume-Of-Fluid (VOF) method to track the interface between the biomass and the surrounding gas phase. Solid-phase pyrolysis reactions are included, and a novel approach is proposed to capture the coupling between the evolution of biomass porosity and the particle shrinkage\tc{, combining different biomass conversion models into one unique framework}. The anisotropic nature of the biomass particle is accounted for in this multidimensional framework. The resulting model \tc{is independent of the number and shape of the particle, and} demonstrates mass conservation and numerical convergence.
Extensive validation with experimental data\tc{, collected from wood particles in the centimetre scale and operating temperature between 400-700\mbox{$^\circ$}C}, shows excellent agreement in terms of mass and temperature profiles and correct volatiles trends. \tc{Predicted char yields fall within 2\% error range}
Shrinking profiles reveal correct trends, \tc{with a 10\% average error in the final particle shape,} but they also highlight the need for a better fundamental understanding of the evolution of the biomass structure. 
Overall, the model takes a step forward in aiding the development of sustainable pyrolysis processes.
The code and simulation setups, developed within the open-source Basilisk framework, are made publicly available.
\end{abstract}


\end{frontmatter}

\section*{Nomenclature}
\subsection*{Acronyms}
\begin{longtable}{p{.15\textwidth}  p{.85\textwidth}}
    VOF & Volume-Of-Fluid\\
    VOS & Volume of Solid\\
    IB & Immersed Boundary\\
    ODE & Ordinary Differential Equation\\
    PLIC & Piecewise Linear Interface Calculation\\
\end{longtable}
\subsection*{Greek letters}
\begin{longtable}{p{.15\textwidth}  p{.85\textwidth}}
    $\epsilon$ & porosity [-]\\
    $\rho$ & density [kg m\ts{-3}]\\
    $\dot\Omega$ & chemical formation rate [kg m\ts{-3} s\ts{-1}]\\
    $\omega$ & mass fraction [-]\\
    $\mu$ & viscosity [Pa s]\\
    $\mathbf{\lambda}$ & thermal conductivity [W m\ts{-1} K\ts{-1}]\\
    $\alpha$ & emissivity [-]\\
    $\sigma$ & Stefan–Boltzmann constant [W m\ts{-2} K\ts{-4}]\\
    $\psi$ & potential field of the pseudo-phase velocity [s\ts{-1}]\\
\end{longtable}
\subsection*{Symbols}
\begin{longtable}{p{.15\textwidth}  p{.85\textwidth}}
    $H$ & Heaviside indicator function [-]\\
    $\mathbf{u_g}$ & intrinsic average gas velocity [m s\ts{-1}]\\
    $\mathbf{v}_g$ & seepage gas velocity [m s\ts{-1}]\\
    $\mathbf{j}$ & diffusive fluxes [kg m\ts{-2} s\ts{-1}]\\
    $\mathbf{u}_p$ & shrinking velocity of the porous matrix [m s\ts{-1}]\\
    $\mathbf{u}_s$ & solid matrix velocity not related to shrinkage [m s\ts{-1}]\\
    $f$ & VOF volume fraction [-]\\
    $p$ & pressure [Pa]\\
    $\mathbf{K}$ & permeability [m\ts{2}]\\
    $T$ & temperature [K]\\
    $C_p$ & specific heat [J kg\ts{-1} K\ts{-1}]\\
    $NS$ & number of phase $k$ species\\
    $NSS$ & number of solid species\\
    $\dot Q_R$ & reaction heat [W m\ts{-3} s\ts{-1}]\\
    $\mathcal{D}$ & gas diffusivity [m\ts{2} s\ts{-1}]\\
    $Z$ & share of chemical reactions contributing to shrinking [-]\\
    $N$ & number of cells per domain side length [-]\\
    $R$ & radius [m]\\
    $D$ & diameter [m]\\
    $M$ & total phase mass [kg]\\
    $t$ & time [s]\\
\end{longtable}
\subsection*{Superscripts and subscripts}
\begin{longtable}{p{.15\textwidth}  p{.85\textwidth}}
    $0$ & initial condition\\
    $g$ & quantity referring to the gas, internal or external\\
    $s$ & quantity referring to the solid matrix\\
    $p$ & quantity referring to the pseudo-phase\\
    $e$& quantity referring to the external ambient phase\\
    $i$ & index of the chemical species\\
    $\Gamma$ & interfacial quantity\\
    $k$ & index of the phases within the pseudo-phase, gas or solid\\
    $bulk$ & bulk quantity\\
\end{longtable}

\section{Introduction}\label{sec:introduction} 

The need for renewable energy and sustainable chemicals has been steadily increasing over the past years \cite{IEA2024}. Due to their wide distribution, easy accessibility, and relatively low prices, biomass has emerged as a candidate to tackle these pressing challenges. In particular, pyrolysis conversion processes, i.e. heating without the presence of oxygen, have been extensively studied in the literature \cite{mohan2006pyrolysis}. This process primarily yields three main products: light incondensable gases, a heavier tar fraction that can be converted into bio-oil (a renewable fuel) \cite{guida2020production}, and a carbon-rich solid residue known as bio-char, which has demonstrated promising applications across various fields \cite{biocharreview}. 
Biomass is constituted by a porous polymeric-fibrous structure, which undergoes significant morphological changes during the conversion process.
The degradation of these chains happens in two coupled and competitive ways: an increase in internal porosity, leading to a less dense and more fragile material, and a shrinkage of the chain length, causing deformation of the medium and inducing compressive stresses \cite{BAROUDI2017206stressfibers}. 
These fibres are also the reason for the anisotropic nature of biomass properties, such as its thermal conductivity being higher along the fibres' direction \cite{kollmann2012principles}.

To support the advancement of biomass pyrolysis conversion processes, numerical models have emerged to optimise and enhance the process yield and selectivity.
A comprehensive review of biomass modelling approaches was published by \citet{NAIDU2025107030}.
In this work, the focus is on the fluid dynamics of particle-scale models, i.e. without direct solution of the internal porous structure. 
Existing modelling approaches can be classified into two main categories: the first ones focus on the evolution of the solid and its internal structure, whereas the second types describe the surrounding environment and the evolution of the released volatiles. 
Starting from the first category, \citet{DIBLASI19961121} developed a 1D model for the shrinking of a biomass particle subjected to a given radiation heat. Coupling simple kinetics with heat and mass transfer phenomena, the authors were able to estimate the released products and correlate the shrinkage to the conversion progress. 
\citet{gentile2017bioSMOKE} developed a 3D anisotropic framework, accounting for detailed wood reaction chemistry using a semi-detailed reaction mechanism involving over fifty species. This model utilised a mobile mesh strategy to account for arbitrary deformation and particle shrinkage during thermal decomposition.
\citet{anca2017online} developed both an experimental apparatus and a numerical model to analyse the released volatile species from the pyrolysis of a cylindrical wood pellet. The authors also considered secondary gas-phase reactions, which further decompose the released volatiles into smaller molecules, obtaining better agreement with the online recorded product profiles. 
More recently, \citet{Pyromech} developed, within the PATO framework \cite{PATO2014}, a model able to correlate the deformation of the wood particle to the mechanical stress arising due to the conversion progress and thermal expansion. The pyromechanical approach, given its physical foundation, effectively captures stress distributions throughout the particle and provides a foundation for future investigations into crack formation and propagation during biomass pyrolysis.
All these models can accurately describe the deformation of a biomass particle, but they neglect the solution of the surrounding environment and, therefore, rely on sub-grid-scale correlations for the estimates of heat and mass transfer coefficients at the boundaries.

Considering the gas-phase models, different approaches that resolve the external environment, neglecting the particle's evolution, have also been proposed. 
These works focus on solid materials which can be described using simpler kinetics, such as coal and metals. 
For example, \citet{tufano2019char} simulated the combustion of a coal particle, achieving good results in both ignition delay times and characteristic degradation times.
Similarly, \citet{hasse2024ironcomb} simulated the combustion of multiple iron particles in an oxidative flow environment, highlighting how particle spacing and diameter influence the transition between continuous and discrete flame propagation modes, with implications for understanding heterogeneous combustion dynamics in metal particle systems.

All the previously mentioned works fall under the same critical assumption: the solid particle interface is considered stationary during the simulation. This hypothesis significantly simplifies the model's complexity but limits its applicability to conditions where this effect can be neglected. For example, the variation of the characteristic dimension, such as the particle diameter, can significantly impact the exchange of heat and mass across the interface \cite{huang2014modeling}. Therefore, it is of crucial relevance to find numerical solutions to consider the shrinking phenomena, which can shed light on the interplay between biomass degradation and volume changes.
In the context of wood pyrolysis, \citet{GOMEZ2018671} combined this effect in a simplified manner, using a mass redistribution approach, without a systematic description of the multiphase system.   

To overcome the limitations of the above approaches, this work proposes a unified framework which can resolve the evolution of the solid particle, together with the dynamics of the gas phase. This model resolves the boundary layer at the gas-solid interface and does not rely on sub-grid-scale correlations for heat and mass transfer. The particle evolves considering both porosity changes and the volume-shrinking effects. The resulting formulation is \tc{\sout{general} is independent of the particle shape and number}, and it can be applied to conditions where the interface exchange terms are not known or cannot be easily estimated, which was not possible with the previous models.

This model employs a single grid to describe both the solid biomass and the gas phase, using an Eulerian approach. In particular, the two-phase system is described using a Volume-Of-Fluid (VOF) model \cite{HIRT1981Vof}, which \tc{\sout{guarantees} has been demonstrated to offer} a sharp and conservative transport of the interface \cite{WEYMOUTH20102853}. The VOF methodology has been employed for simulations of several physical phenomena, such as wave breaking \cite{Deike2015wavebreaking}, pulsed jets \cite{Kulkarni2025jet}, droplet coalescence and breakups \cite{Wang2025dropletcollision, Farsoiya2023dropletbreakup}, bubbly flows \cite{MULBAH2022bubblyflow}, droplet evaporation and combustion \cite{CIPRIANO1,CIPRIANO2, CIPRIANO3}. 
\tc{Other works have used VOF-like indicator functions to track evolving fluid-solid interfaces. For example, }\citet{maes2022improved} \tc{\sout{already applied the VOF to describe the flow within porous media, focusing on the description of pore-scale phenomena.} applied a "Volume-of-Solid" (VOS) approach at the pore scale within a micro-continuum framework, where the volume fraction field captures the evolving solid-fluid boundary during mineral dissolution. In our work, the VOF indicator tracks the interface between the external gas and a porous pseudo-phase with significant internal flow. At the same time, a separate porosity field describes the internal structure. In this work, the term "Volume-of-Fluid" is preferred as all numerical methods employed derive directly from the established VOF literature.}
\tc{Furthermore, \sout{In this work,}} the emphasis is \tc{put} on the larger particle scale, extending the VOF approach to simulate the evolution of a reacting porous particle \tc{rather than focusing on the evolution of the individual pores}. This is made possible through several advancements: i) incorporating porosity within the particle; ii) introducing Darcy–Forchheimer terms into the momentum equations to account for drag effects; iii) modelling chemical reactions in the solid phase; iv) developing an accurate framework to handle porosity variations and particle shrinkage while ensuring mass conservation. This approach \tc{\sout{eliminates} limits} the numerical instabilities and mass conservation issues typically encountered when using separate grids to represent the solid particle and the surrounding gas phase.
The governing equations for this model are described in \autoref{sec:mathematical-formulation}, together with the underlying assumptions and the biomass-gas boundary conditions. The numerical methods for solving these equations are summarised in \autoref{sec:numerical-strategy}. After a brief verification in \autoref{sec:numerical-tests}, the model is used to simulate realistic biomass particles (\autoref{sec:applications}), which are compared with experimental data in terms of mass and volume changes. The model captures the \tc{\sout{correct} main} experimental trends and \tc{\sout{explains} it takes a step forward in the understanding of} the distribution of shrinking and porosity changes inside the particle, which are difficult to measure experimentally.

\tc{The proposed model is independent of the particle shape and number, and provides an excellent candidate for the discovery of sub-scale correlation useful for reactor scale frameworks, like bubbling bed or auger reactors. In particular, models for larger-scale systems are unable to resolve the interface between the biomass and the environment, as well as the evolution of the physical quantities within and around the particles. This information is usually based on literature sub-grid-scale correlations that must be introduced into large-scale models. However, an accurate description of the system requires models tailored for the system and for the operative conditions being solved. In light of this, the present model allows refining the understanding of the particle behaviour to obtain specific correlations for simulating industrial-scale reactors following a multi-scale strategy.}

\section{Mathematical Formulation}\label{sec:mathematical-formulation}
Capturing the evolution of biomass pyrolysis at the particle-scale requires the introduction of control volumes which are much larger than the pore size and much smaller than the particle’s characteristic dimension. 
\begin{figure*}
	\centering
    \usetikzlibrary{intersections}
\usetikzlibrary{arrows.meta}
\usetikzlibrary{calc}
\usetikzlibrary{shapes.arrows}
\usetikzlibrary{patterns} %
\usetikzlibrary{patterns, patterns.meta, decorations.pathmorphing}
\definecolor{Lightgreen}{HTML}{87A896}
\definecolor{Darkgreen}{HTML}{004225}

\usetikzlibrary{patterns}
\pgfdeclarepatternformonly{big dots}{\pgfqpoint{-1pt}{-1pt}}{\pgfqpoint{4pt}{4pt}}{\pgfqpoint{4pt}{4pt}}{
    \pgfpathcircle{\pgfqpoint{0pt}{0pt}}{1pt}
    \pgfusepath{fill}
}

\begin{tikzpicture}[scale=2]
    \def\xmin{0}
    \def\xmax{3.5}
    \def\ymin{0}
    \def\ymax{3.5}
    
    \draw[thick, fill=Lightgreen] 
        (\xmin,\ymin) -- 
        (\xmin,1.8) .. controls (0.5,2.5) and (1,2.2) .. 
        (1.5,1.8) .. controls (2,1.4) and (2.5,1.2) .. 
        (\xmax,1.2) -- 
        (\xmax,\ymin) -- 
        cycle;

    \draw[thick, pattern=big dots, pattern color=white] 
        (\xmin,\ymin) -- 
        (\xmin,1.8) .. controls (0.5,2.5) and (1,2.2) .. 
        (1.5,1.8) .. controls (2,1.4) and (2.5,1.2) .. 
        (\xmax,1.2) -- 
        (\xmax,\ymin) -- 
        cycle;

    \draw [->, >=Triangle, thick] (1.665, 1.665) -- (1.97,2.1);
    
    \node at (2.1,1.9) {$\mathbf{n}_\Gamma$};
    \node at (0.2,2.2) {$\Gamma$};
    
    \draw[thick] (\xmin,\ymin) rectangle (\xmax,\ymax);
    
    \node[draw=black] at (2.6,3.2) {gas phase $H=0$};
    \node[fill=gray!20!white, draw=white, text=black, inner sep=2pt] at (1,0.3) {pseudo phase $H=1$};

    \coordinate (C1) at (2.6,0.6);
    \coordinate (C2) at (5,1.75);
    \def\r{0.2}
    \def\R{1.2}
    
    \draw[thick] (C1) circle (\r);
    \fill[thick, fill=white] (C2) circle (\R);
    
    \pgfmathsetmacro{\d}{sqrt((5-2.6)^2 + (1.75-0.6)^2)}
    
    \pgfmathsetmacro{\alpha}{asin((\R-\r)/\d)}
    \pgfmathsetmacro{\beta}{atan2(1.75-0.6,5-2.6)}
    
    \pgfmathsetmacro{\gamma}{asin((\R+\r)/\d)}
    
    \coordinate (T1a) at ($(C1) + (\beta+\alpha+90:\r)$);
    \coordinate (T1b) at ($(C1) + (\beta-\alpha-90:\r)$);
    \coordinate (T2a) at ($(C2) + (\beta+\alpha+90:\R)$);
    \coordinate (T2b) at ($(C2) + (\beta-\alpha-90:\R)$);
    
    \coordinate (T1c) at ($(C1) + (\beta+\gamma+90:\r)$);
    \coordinate (T1d) at ($(C1) + (\beta-\gamma-90:\r)$);
    
    \draw[thick] (T1a) -- (T2a);
    \draw[thick] (T1b) -- (T2b);

\begin{scope}
        \clip (C2) circle (\R);
        \filldraw[Lightgreen, draw=black] 
    plot[smooth cycle, tension=1.2,] coordinates {
        (5, 1.5) (5.2, 1.8) (5.3, 1.4) (4.9, 1.3) (4.7, 1.4)
    };
        \filldraw[Lightgreen, draw=black] 
    plot[smooth cycle, tension=1.1,] coordinates {
        (5.7, 2.1) (6.0, 2.3) (5.9, 1.9) (5.5, 1.8) (5.4, 2.0)
    };
    \filldraw[Lightgreen, draw=black] 
    plot[smooth cycle, tension=1.3,] coordinates {
        (4, 1.8) (3.5, 2) (4.3, 2.1) (4.6, 1.6) (4.3, 1.5)
    };
    
    \filldraw[Lightgreen, draw=black] 
    plot[smooth cycle, tension=1.0,] coordinates {
        (4.9, 0.5) (5.2, 0.6) (5.4, 0.8) (5.1, 1.0) (4.7, 0.9)
    };
    
    \filldraw[Lightgreen, draw=black] 
    plot[smooth cycle, tension=1.0,] coordinates {
        (4.2, 0.7) (4.4, 0.8) (4.5, 1) (4.2, 1.2) (4., 1.1)
    };
    
    \filldraw[Lightgreen, draw=black] 
    plot[smooth cycle, tension=1.4,] coordinates {
        (4.8, 3) (5.1, 2.8) (5.3, 2.6) (5.0, 2.1) (4.6, 2.2)
    };
    
    \filldraw[Lightgreen, draw=black] 
    plot[smooth cycle, tension=1.2,] coordinates {
        (5.8, 1.7) (6.3, 1.7) (6.1, 1.4) (5.8, 1.5) (5.5, 1.5)
    };
    
    \filldraw[Lightgreen, draw=black] 
    plot[smooth cycle, tension=1.3,] coordinates {
        (5.6, 0.9) (5.9, 0.8) (6.0, 1.1) (5.7, 1.3) (5.4, 1.1)
    };
\end{scope}

    \draw[thick] (C2) circle (\R);

    \draw[thick] (4.8, 2.3)  -- (4.3, 3.1) node [above] {Solid};    
    \draw[thick] (5, 1.2)  -- (5.3, .3) node [below] {Internal gas};    
\end{tikzpicture}
    \caption{Control volume encapsulates both the external environment and the pseudo-phase with interface $\Gamma$ and interface normal $\mathbf{n}_\Gamma$}
	\label{fig:control-volume}
\end{figure*}
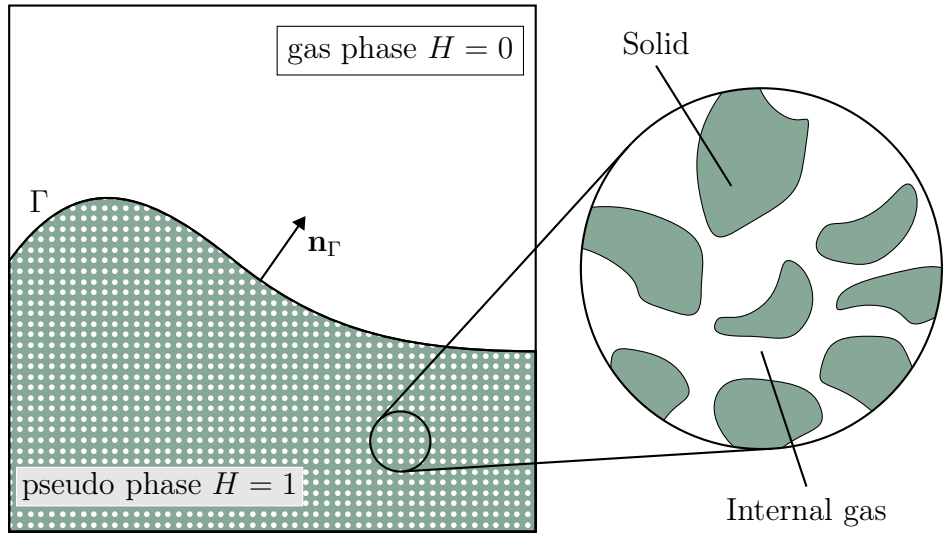
\autoref{fig:control-volume} displays the sharp distinction between the gas phase and the \tc{biomass-}particle phase. The latter is treated as a single pseudo-phase as it contains a mixture of solid matrix and gas phase within the pores. Mathematically, we distinguish between the \tc{\sout{two} different} phases by introducing a sharp and a smooth indicator.
The sharp indicator function $H$ differentiates between the external gas phase and the porous pseudo-phase, and it is defined as:
\begin{equation}\label{eq:indicator-definition}
    H(\mathbf{x}, t) =
    \begin{cases}
        1 & \text{if pseudo-phase $(p)$} \\
        0 & \text{if environment gas phase $(e)$}
    \end{cases}
\end{equation}
Following this definition, the interface between the gas and the particles is transported through the solution of the transport equation \cite{drew1982mathematical}:
\begin{equation}\label{eq:indicator-transport}
    \dfrac{\partial H}{\partial t} + \mathbf{u}_p \cdot \nabla H = 0
\end{equation}
The velocity $\mathbf{u}_p$ is the pseudo-phase shrinking velocity, which depends on the chemical reactions occurring within the pseudo-phase. Its calculation is explained in detail in \autoref{sec:shrinking-velocity}.
While sharp-interface multiphase models usually adopt the indicator function \cite{tryggvason2011direct}, the peculiarity of this work is the presence of the non-homogeneous pseudo-phase, described by introducing a porosity field $\epsilon_g$, which is defined as the volume fraction of gas in each control volume $V$:
\begin{equation}\label{eq:porosity-definition}
    \epsilon_g(\mathbf{x},t) = \frac{V_g}{V} = 1-\epsilon_s(\mathbf{x},t)
\end{equation}
This definition implies that the porosity has fractional values within the porous particle, while it is identically equal to $1$ in the gas phase environment. 
Assuming that the control volume is much larger than the pore size, the porosity results in a smooth function, in analogy to the formulation employed by two-fluid models \cite{marschall2011towards}. \tc{The pseudo-phase is treated as a continuum, where overall phase properties are calculated as an average weighted on the porosity scalar field.}
The evolution of the porosity is affected by the chemical reactions and the velocity of the pseudo-phase, as discussed in the next sections.
Given that the calculation of the porosity is computed by volume averaging over a sufficiently large dimension with respect to the pore size, any random cross-section is statistically representative of the entire 3D structure. Therefore, the volumetric porosity is also equal to the surface porosity, which is used in the calculation of fluxes across the domain \cite{ConvectionPorousMedia}.
\subsection{Governing equations}\label{sec:pseudo-phase}

The equations describing the evolution of the pseudo-phase and the external environment are derived from a set of conservation laws governing mass, momentum, energy, and chemical species. 
We introduce the subscript $k$, which is used to distinguish between the gas and the solid phase, considering that $k=s$ indicates solid phase properties, while $k=g$ denotes the gas phase, either in the external environment or within the pseudo-phase.
Following \citet{hsu1990thermal}, and including the effect of the particle’s shrinking velocity $\mathbf{u}_p$, the following system of equations is obtained:
\begin{equation}\label{eq:totphasemass}
\frac{\partial \left(\epsilon_k\rho_k\right)}{\partial t} +\nabla\cdot\left(\rho_k\epsilon_k(\mathbf{u}_k-\mathbf{u}_p)\right)=  \sum^{NS}_{i=1}\dot\Omega_{i}
\end{equation}
\begin{equation}\label{eq:gasspecies}
\frac{\partial \left(\epsilon_k\rho_k\omega_{k,i}\right)}{\partial t} +\nabla\cdot\left(\rho_k\epsilon_k\omega_{k,i}(\mathbf{u}_k-\mathbf{u}_p)\right)= -\nabla\cdot(\epsilon_k\mathbf{j}_{k,i})+\dot\Omega_{i} 
\end{equation}
\begin{equation}\label{eq:DBS}
    \frac{\partial (\rho_g\epsilon_g\mathbf{u}_g)}{\partial t}+\nabla\cdot\left(\rho_g\epsilon_g\mathbf{u}_g \mathbf{u}_g\right) = -\nabla\tc{\cdot} (\epsilon_g\mathbf{T}) + \mathbf{F}_{drag}H
\end{equation}
\begin{equation}\label{eq:pf-energy-eq}
        \epsilon_k \left( \rho C_p\right)_k \frac{\partial T_k}{\partial t} +  \left( \rho C_p \right)_k \epsilon_k( \mathbf{u}_k-\mathbf{u}_p)\cdot\nabla T_k = \nabla\cdot\left( \epsilon_k\boldsymbol{\lambda}_k \nabla T_k\right) - \nabla T_k\cdot \left(\sum_{i=1}^{NS}{C_p}_i\epsilon_k\mathbf{j}_{k,i} \right) +\dot Q_R
\end{equation}
with $\mathbf{T}=\mu_k(\nabla\mathbf{u}_k +\nabla\mathbf{u}_k^T) - p\mathbf{I}$.

For $\epsilon_g\rightarrow1$, this set of equation tends to the standard set of conservation laws for multicomponent single-phase systems.

The following considerations/assumptions can then be made: 
\begin{itemize}
    \item Ideal gas, Newtonian flow behaviour, and low-Mach number are assumed for the gas both inside the pseudo-phase and in the surrounding environment.
    \item In the pseudo-phase, gas is allowed to flow within the solid matrix in the interconnected region of the pores. Two different gas average velocities can be identified, depending on the control volume chosen during the averaging operation. Considering only the gas volume $V_g$, the intrinsic average velocity $\mathbf{u}_g$ is recovered, which is the velocity the molecules of gas experience moving within the pores. Considering the whole volume of the pseudo-phase $V$, the Darcy or seepage velocity $\mathbf{v}_g$ is obtained instead. The two are related through the Dupuit-Forchheimer relationship \cite{ConvectionPorousMedia}:
    \begin{equation}\label{eq:df-relashionship}
        \mathbf{v}_g = \epsilon_g\mathbf{u}_g
    \end{equation}
    The current system of equations for the gas is presented in terms of $\mathbf{u}_g$, but it is solved in terms of the Darcy velocity, given its simpler implementation and intuitiveness. 
    \item \autoref{eq:DBS} is solved only for the gaseous phase. For all the analysed cases, $\mathbf{u}_s$ (i.e. the solid velocity without the shrinking contribution $\mathbf{u}_p$) is considered null as all the experimental measurements reported in \autoref{sec:applications} used stationary particles, fixed in position after being inserted in the pyrolysis environment. From another perspective, this is equivalent to working in a reference frame that moves with the particle, so that the particle's velocity in this frame is always zero.
    \item The $\mathbf{F}_{drag}$ terms represent the viscous drag force experienced by the fluid moving within the porous matrix. It is non-zero only within the pseudo-phase and can be expressed as:
\begin{equation}\label{eq:darcy-forchheimer}
    \mathbf{F}_{drag} = - \left[ \frac{\mu_g \epsilon_g \mathbf{v}_g}{\mathbf{K}} + \rho_g\frac{1.75}{\sqrt{150\epsilon_g^3}} \frac{\epsilon_g|\mathbf{v}_g|\mathbf{v}_g}{\sqrt{\mathbf{K}}} \right]
\end{equation}
    The two terms represent, respectively, the Darcy and Forchheimer contributions, and $\mathbf{K}$ is the medium permeability in m\ts{2}.
    \item Solid-phase chemical reactions can occur throughout the entire pseudo-phase domain, not just at the interface with the ambient phase. Degrading solid species can either release volatiles or transform into different solid components. $\dot\Omega_{i}$ represents the net chemical formation rate for species $i$ and is calculated per unit volume. Reactions can also occur in the internal and surrounding gas phase, where the produced species can further decompose into smaller products \cite{ANCACOUCE2017411} or grow forward into carbonaceous nanoparticles \cite{SERSE2023100263}. The model can already consider purely gas-phase reactions, but given the low temperatures and high dilutions at which the pyrolyses are conducted, these reactions are neglected in this work. When dealing with gasification or combustion conditions, this hypothesis is no longer valid, and the gas-phase reactions must be properly accounted for. The application of the proposed model in these conditions will be the subject of future studies.
    \item The diffusive fluxes $\mathbf{j}_{k,i}$ are considered only for gas species and are calculated using a mixture-averaged approach \cite{Cuoci2013laminarsmoke}. Within the pseudo-phase, the presence of the solid matrix hinders diffusion. This effect is considered by correcting the pure gas phase diffusion coefficient using the porosity \cite{ConvectionPorousMedia, millington1961permeability}, such that $ \mathcal{D}_{p,i}=\mathcal{D}_{e,i}\space \epsilon_g^{1/3}$.
    \item Soret effects is neglected as we do not consider small molecules which exhibit strong thermophoretic effects. 
    \item Thermal equilibrium is assumed between the internal gas and the solid \cite{DIBLASI19961121}, meaning $T_s=T_g$ within the porous medium. This implies that just one energy equation can be solved for the whole pseudo-phase temperature $T_p$, obtained by summing \autoref{eq:pf-energy-eq} for the internal gas and the solid phase. The pseudo-phase physical properties are computed using an arithmetic average based on the porosity. By doing so, a generic property $\phi$ is calculated as:
\begin{equation}\label{eq:phase-properites}
    \phi_m=\phi_s\epsilon_s+\phi_g\epsilon_g
\end{equation}
    Furthermore, the radiation contribution to the thermal conductivity of the solid matrix $\boldsymbol{\lambda}_m$ is neglected as it is orders of magnitude lower than the purely conductive contribution due to the low temperature analysed. 
    \item To account for the anisotropic nature of biomass, both the solid thermal conductivity $\boldsymbol{\lambda}_s$  and the medium permeability $\mathbf{K}$ can have different values along different directions.
    \item The intrinsic solid density $\rho_s$, also called the skeletal density, is constant in time and space.
\end{itemize}

\subsection{Interface boundary condition}\label{sec:interface-eq}
To derive boundary conditions for the proposed system of equations, a control volume encompassing a portion of the interface $\Gamma$ between the biomass particle and the external environment is considered. Given that the interface has zero thickness, no accumulation can occur in this region. Introducing the jump notation for a generic quantity $\phi$, defined as: $[\phi]_\Gamma = \phi_p - \phi_e$, representing the difference between the value within the porous medium and the external ambient. The interface boundary conditions can then be written as:
\begin{equation}\label{eq:interface-velocity}
    \left[\mathbf{v}_g\right]_\Gamma=[\epsilon_g\mathbf{u}_g]_\Gamma =0
\end{equation}
\begin{equation}\label{eq:interface-pressure}
    \left[p\right]_\Gamma = 0
\end{equation}
\begin{equation}\label{eq:interface-species-balance}
    \left[\mathbf{j}_{g,i}\cdot\mathbf{n}_\Gamma\right]_\Gamma = 0
\end{equation}
\begin{equation}\label{eq:interface-energy-balance}
    \left[\lambda \nabla T \cdot\mathbf{n}_\Gamma \right]_\Gamma = \alpha\sigma\left(T_{g, bulk}^4 - T_\Gamma^4\right) 
\end{equation}
These equations are derived from the conservation of mass, momentum, species and energy, respectively.
 \autoref{eq:interface-velocity} and \autoref{eq:interface-pressure} imply the continuity of gas phase velocity and pressure across the interface.
 The gradients in \autoref{eq:interface-species-balance} and \ref{eq:interface-energy-balance} refer to gradients computed along the interfacial normal direction  $\mathbf{n}_\Gamma$. %
Temperature and species mass fractions are continuous across the interface.
 The following equations are solved numerically to find the interface temperature $T_\Gamma$ and interface gas mass fractions $\omega_{i,\Gamma}$, characterising in detail the interface conditions and the exchange between the porous media and the external gas.

\section{Numerical solution strategy}\label{sec:numerical-strategy}
This section aims to highlight some key steps in the numerical solution of the proposed system of equations.
We chose to implement the proposed model within the open-source code Basilisk \cite{popinet2015quadtree, basiliskwebsite}, because it features an efficient quad/octree grid, which is compatible with VOF and Navier--Stokes solvers.
The proposed model is spatially discretised using the finite volume method, ensuring conservation of the transported quantities \cite{ferziger2002computational}. The numerical solution of the Navier--Stokes equations is based on a time-splitting projection method \cite{chorin1968numerical}, and each scalar transport equation is resolved in a segregated manner, by splitting each term (advection, diffusion, and reaction), in order to discretize them carefully.

A single set of Navier-Stokes equations is solved for the gas velocity in both phases, using a \textit{one-field} approach \cite{tryggvason2011direct}.\tc{\sout{Details on the solution algorithm for these equations can be found in the work} The solution algorithm is based on a second-order (in space and time) approximate projection method \cite{POPINET2003572, BCG}, which was developed for incompressible two-phase flows} by \citet{POPINET20095838} \tc{and extended to include volumetric sources. The latter makes the velocity field non-solenoidal, thereby making the convergence rate for the particle expansion/shrinkage first-order accurate.} 
This algorithm is \tc{further} modified in this work in order to include the physical phenomena which characterise the porous systems in this work, as described in the following sections.
All the face values in the convective terms are discretised using the second-order upwind method from \citet{BCG}. Interface transport is performed using a directional split \tc{first-order in time forward-Euler} approach \cite{scardovelli1999direct} in the conservative form proposed by \citet{WEYMOUTH20102853}. This method calculates the convective fluxes through a geometric reconstruction of the interface. The same fluxes are also used to solve the convective transport of the scalar fields (e.g. temperature, species mass fractions), which are effectively treated as \emph{tracers} of the VOF fraction \cite{LOPEZHERRERA201514}.
Furthermore, the solution of the temperature and species field is carried out using a \textit{two-field} approach. This means that two different governing equations are solved on either side of the interface, with exchange contributions between them. This approach limits unwanted numerical diffusion in the other phase, allows for a precise interface solution, and was found to be substantially more effective when dealing with chemical reactions. 
Precise temperature and species concentrations are essential for accurate reaction rate calculations, as the exponential temperature dependence can magnify even small variations. Exchange terms between the two fields are found by solving the interface balance equations listed in \autoref{sec:interface-eq} and are introduced as an explicit source term during the scalar diffusion step. This procedure is equivalent to setting Neumann boundary conditions for the particle-environment interface, similarly to what is done in previous works \cite{CIPRIANO1,zhao2022boiling}.

\subsection{Darcy and Forchheimer contributions}\label{sec:darcyterm}
The Darcy and Forchheimer terms are directly included within the solution of the Navier-Stokes equations, and they are non-zero only within the pseudo-phase.
Given the very low values of the wood permeability vector $\mathbf{K}$ (commonly in the order of $10^{-14}$ m\ts{2}), it is essential that these terms are treated either fully or partially implicitly, as an explicit formulation would require unfeasibly low time steps for stability. 
Using an operator-splitting approach \tc{it is possible to isolate the effect of the drag term in \autoref{eq:DBS}, separating it from the rest of the terms of the Navier--Stokes equation. Noticing that the \mbox{$\mathbf{F}_{drag}$} term is multiplied by the indicator function \mbox{$H$}, we can use a volumetric average operation to simplify \autoref{eq:DBS} to \sout{ and volumetric averaging, \mbox{\autoref{eq:DBS}} can be simplified to}}:
\begin{equation}\label{eq:darcy-discretisation}
  \rho_g\frac{\partial \mathbf{v}_g}{\partial t} = - \left[A \mathbf{v}_g +  B|\mathbf{v}_g|\mathbf{v}_g\right]f
\end{equation}
with $A=\frac{\mu_g \epsilon_g}{\mathbf{K}}$, $B=\rho_g\frac{1.75}{\sqrt{150\epsilon_g^3}} \frac{\epsilon_g}{\sqrt{\mathbf{K}}}$\tc{, \sout{and}} $f$ is the VOF \tc{pseudo-phase} volume fraction defined as $f=\frac{1}{V}\int_VHdv$\tc{, and \mbox{$\rho_g$} and \mbox{$\mu_g$} are respectively the intrinsic internal gas-phase density and viscosity. In the interfacial cells, $\mathbf{F}_{drag}$ only acts on the biomass portion of these cells. Since it is defined per unit volume of biomass, multiplying by $f$ converts it to a force per unit volume of the entire cell.}
The contribution of the Darcy and Forchheimer terms is applied after the predictor step in the Navier-Stokes solution algorithm.
Consider $\mathbf{v}_g^*$ the intermediate gas velocity field found after the predictor step and $\mathbf{v}_g^{**}$ the new velocity field after this correction step.
\autoref{eq:darcy-discretisation} is discretised in a semi-implicit way, linearising the Forchheimer contribution using the magnitude of $\mathbf{v}_g^*$. Doing so, integrating between $t$ and $t+\Delta t$, the gas velocity can be updated as:
\begin{equation}\label{eq:darcy-velocity-update}
    \mathbf{v}_g^{**} = \mathbf{v}_g^*exp\left(-\frac{A+B|\mathbf{v}_g^* |}{\rho_g}\Delta t \space f\right)
\end{equation}
This semi-implicit formulation is stable, as the Forchheimer term is usually one to two orders of magnitude lower than the Darcy contribution in the analysed conditions. 

\tc{A validation case for the implementation of this term is presented in \autoref{sec:BaJ-problem}}

\subsection{Chemical reaction step}
The reaction terms are introduced by solving a batch-reactor-like equation for each cell, updating the values of the solid and gas mass fractions according to:
\begin{equation}\label{eq:batch}
    \frac{\partial(\epsilon_k\rho_k\omega_{k,i})}{\partial t} = \dot\Omega_{i}
\end{equation}
Two more equations are also solved to update the total porosity (see also \autoref{sec:shrinking-velocity}) and the phase temperature. 
The total resulting system of ODE is tightly coupled due to the complex reaction network. Therefore, a stiff ODE solver is necessary, as non-A-stable integration of this step could lead to unfeasible values for the mass fractions. In this framework, this is achieved by exploiting the already implemented solvers within the OpenSMOKE++ library \cite{CUOCI2015237}. Efficient chemistry integration is also essential as the computational effort needed for this step increases quadratically with the number of species involved, often resulting in a bottleneck for the time-stepping algorithm.

\subsection{Chemical kinetics and material properties}
The model employs a CHEMKIN format \cite{kee1996chemkin}, managed through the OpenSMOKE++ library \cite{CUOCI2015237}, to store and read arbitrarily complex kinetic mechanisms. These mechanisms are used to compute formation rates as well as the required thermodynamic and transport properties, which are evaluated at each time step. In this work, the most recent biomass mechanism developed by CRECK, including the updated cellulose submodel from \citet{debiagi2024cellulose}, is adopted.
The complete kinetic scheme, together with the corresponding transport and physical property data, is provided as supplementary material.
Because of their strong variability, the solid-phase density ($\rho_s$), thermal conductivity ($\boldsymbol{\lambda}_s$), and initial porosity ($\epsilon_g^0$) were determined separately for each case based on experimental measurements. \tc{The choice of these parameters significantly affects the model output. Literature correlation or direct measurements are preferred whenever possible to maintain a general formulation.} The initial biomass surrogate composition was estimated from the C/H/O ultimate analysis of the investigated sample, following the procedure of \citet{debiagi2015extractives}. \tc{Permeability measurements are often not available/performed in the same study where pyrolysis data are collected. Therefore, average reported values for the same wood type were used. Moreover, based on the authors' experience, permeabilities \mbox{$\mathbf{K<}1\times10^{-8} $} showed no significant difference, as this low value already implies complete blockage to the external flow. The minor differences are attributable to the internal pressure reached, which has only a minor effect on the variation of physical properties compared to temperature.}

\subsection{Calculation of the shrinking velocity}\label{sec:shrinking-velocity}
Considering a stationary solid phase ($\mathbf{u}_s=0$) subject only to a shrinking velocity $\mathbf{u}_p$. \autoref{eq:totphasemass} can be rewritten as:
\begin{equation}\label{eq:totsolidmass}
\frac{\partial \epsilon_s}{\partial t} + \mathbf{u}_p\cdot\nabla\epsilon_s + \epsilon_s\nabla\cdot\mathbf{u}_p=  \frac{1}{\rho_s}\sum^{NSS}_{i=1}\dot\Omega_{i}
\end{equation}
The first two terms on the left-hand side represent the porosity change, whereas the third term represents the volume variation. An additional constraint is necessary to solve this one equation in the two unknowns $\epsilon_s$ and $\mathbf{u}_p$. Consider the right-hand side term equal to a generic source term S to simplify the notation, so that $S =\frac{1}{\rho_s}\sum^{NSS}_{i=1}\dot\Omega_{i} $. To saturate the missing degree-of-freedom, the reactive source term is split into two contributions according to an arbitrary function $Z$, so that:
\begin{equation}\label{eq:porosity-variation}
    \frac{\partial \epsilon_s}{\partial t} + \mathbf{u}_p\nabla\epsilon_s =  S (1-Z)
\end{equation}
\begin{equation}\label{eq:volume-variation}
 \epsilon_s\nabla\cdot\mathbf{u}_p=  S Z
\end{equation}
Summing \autoref{eq:porosity-variation} and \ref{eq:volume-variation} results back to \autoref{eq:totsolidmass}. $Z(\mathbf{x},t)$ is an arbitrary function of space and time, bounded between $0$ and $1$, that distributes the reaction term between volume variation and porosity change.
$Z = 1$ corresponds to a pure shrinking condition, with no porosity change. On the contrary, $Z = 0$ indicates a fixed interface,  with only a variation in internal porosity.
\tc{Given that it is not clear what causes and how the shrinking phenomena occur, the proper expression for this function remains unknown.}
In principle, the functional form of $Z$ should be related to the physics of the shrinking problem. For example, one could develop an expression correlating the solid properties (i.e., temperature, pressure, and composition) with the importance of shrinking compared to the porosity changes. The functional form of $Z$ serves as a closure model required to solve \autoref{eq:totsolidmass} and can also be discontinuous. In this work, various forms of this function are investigated, with a detailed discussion provided in \autoref{sec:mass-conservation}. Furthermore, a general expression is proposed and tested against real experimental data in \autoref{sec:applications}.
\autoref{eq:volume-variation} can be solved using a potential flow approximation. The velocity $\mathbf{u}_p$ can be set equal to the gradient of a potential field $\psi$:
\begin{equation}\label{eq:velocity-potential}
    \mathbf{u}_p = -\nabla\psi
\end{equation}
Introducing \autoref{eq:velocity-potential} in \autoref{eq:volume-variation}, a Poisson equation is obtained. This equation is solved numerically to find the potential field $\psi$ \cite{popinet2015quadtree}.
\begin{equation}\label{eq:poisson-velocity-potential}
     \nabla^2\psi=  \frac{S Z}{\epsilon_s}
\end{equation} 
Once $\psi$ is known, the calculation of the regression velocity is straightforward, and the interface movement is computed according to \autoref{eq:indicator-transport}. 
\subsection{Divergence of the gas phase}\label{sec:gas-divergence}
As previously mentioned, the gas phase is solved using a one-field approach. It is therefore useful to reformulate \autoref{eq:totphasemass} in a more convenient form to express the gas phase velocity divergence. This can be done considering \autoref{eq:totphasemass} written for the internal gas phase, external gas phase and solid mass. Summing these three equations leads to: 
\begin{equation}\label{eq:gas-divergence}
    \nabla\cdot\mathbf{v}_g= \epsilon_s\sum_{i=1}^{NSS}\dot\Omega_i\left[\frac{1}{\rho_g}-\frac{1}{\rho_s}\right]H - \epsilon_g\frac{1}{\rho_g}\frac{D\rho_g}{Dt}
\end{equation}
\autoref{eq:gas-divergence} highlights that the expansion of the gas is due to two different contributions: the first one is related to the production of gas due to solid/biomass chemical reactions, whereas the second one is the variation in the gas phase's density due to pressure and temperature changes. Written in this form, the divergence of the gas phase can be easily used to overload the projection step of the Navier-Stokes solution \cite{CIPRIANO1, POPINET2003572}, finding a flow field satisfying a non-zero divergence condition.

\subsection{Time-stepping algorithm}
For completeness, the whole time-stepping algorithm is reported in \autoref{alg:one}. After an initialisation step, where allocation of the necessary memory for the various fields is performed, the time loop begins. 

\tc{The time step is computed such that it satisfies a Courant-Friedrich-Lewis condition of 0.5, which guarantees the stability of the directional split transport of the VOF scheme \cite{WEYMOUTH20102853}. Simulation with an initial surrounding stationary gas phase would imply an unfeasibly large time step, leading to unstable conditions. Therefore, a maximum time step must be set to ensure stability of the temperature and species equations at the initial time step, when the velocity is zero.}

At each time step, we first compute the chemistry integration to calculate the interface velocity $\mathbf{u}_p$. Once this is established, movement of the interface and the associated tracer fields is performed. Recalculation of the physical properties must now be carried out, as some computational cells could have been emptied of the pseudo-phase tracer or, vice versa, could now include a portion of it. After this step, we proceed with the solution of the Navier-Stokes equation to find the gas velocity $\mathbf{v}_g$. The gravity effect can also be included as an additional acceleration term during this part of the algorithm. Basilisk also has a strong adaptive mesh refinement algorithm, which can be optionally used to reduce the total number of computational cells and, therefore, the computational time. 
\tc{The grid is refined with a quad/octree strategy, using a wavelet-based error estimation to avoid issues related to the numerical gradients calculation. This approach is described by \mbox{\citet{van2018towards}}. The target fields for adaptation of the simulations in this work are: temperature, velocity, inert gas mass fraction and the biomass main released product volatile fraction (levoglucosan). The refinement is adjusted so that the interface always remains at the maximum level of refinement, to guarantee the correct calculation of the interfacial gradients for the solution of the species and thermal boundary layer.}

After logging the key interest variables, the iteration is repeated until the final time is reached.

{\linespread{1}\selectfont
\begin{algorithm}
\DontPrintSemicolon
\SetKwBlock{KwNS}{\textcolor{ForestGreen}{$\rhd$ \textbf{\emph{Navier-Stokes solution, calculation of $\mathbf{v}_g$}}}}

\SetKwBlock{KwInit}{\textcolor{ForestGreen}{$\rhd$ \textbf{\emph{Initialisation}}}}

\SetKwBlock{KwChem}{\textcolor{ForestGreen}{$\rhd$ \textbf{\emph{Chemistry}}}}

\SetKwBlock{KwInterface}{\textcolor{ForestGreen}{$\rhd$ \textbf{\emph{Interface and tracers}}}}
\Begin{
    \KwInit{
        $i \gets 0$, $t \gets 0$\; 
        Kinetic scheme loading\;
        Fields allocation and initialisation \;
        Initial physical properties calculation\;
    }
    \While{$t<t_{end}$}{
        $i\gets i+1$\;
        Stable $\Delta t$ calculation\; 
        \KwChem{
            $\dot\Omega_{i}$ calculation and chemistry integration (\myeqref{eq:batch}, \ref{eq:porosity-variation})\;
            Update of $Z$ field\;
            Calculation of shrinking velocity $\mathbf{u}_p$ (\myeqref{eq:velocity-potential}, \ref{eq:poisson-velocity-potential})\;
        }
        \KwInterface{
            Interface movement (\myeqref{eq:indicator-transport})\;
            Tracers advection\;
            Interface solution and exchange terms calculation\;
            Tracers diffusion\;
        }
        Physical properties update\; 
        \KwNS{
            Predictor step\;
            Darcy-Forchheimer contribution (\myeqref{eq:darcy-velocity-update})\;
            Viscous term\;
            Additional acceleration terms\;
            Projection step (\myeqref{eq:gas-divergence})\;
        }
        Mesh adaptation\;
        $t \gets t+\Delta t$\;
        Output variables
    }
    Field deallocation and memory clean-up\;
}
\caption{Time-stepping algorithm}\label{alg:one}
\end{algorithm}
}

\section{\tc{Numerical tests}}\label{sec:numerical-tests}
\subsection{\tc{Beavers and Joseph Problem}}\label{sec:BaJ-problem}
\tc{ We first want to test the proposed formulation in a non-reactive case, to verify the correct implementation of the Darcy drag term.}   

\begin{figure*}
    \centering
        \begin{subfigure}{0.48\textwidth}
        \centering
        \scalebox{0.8}{\input{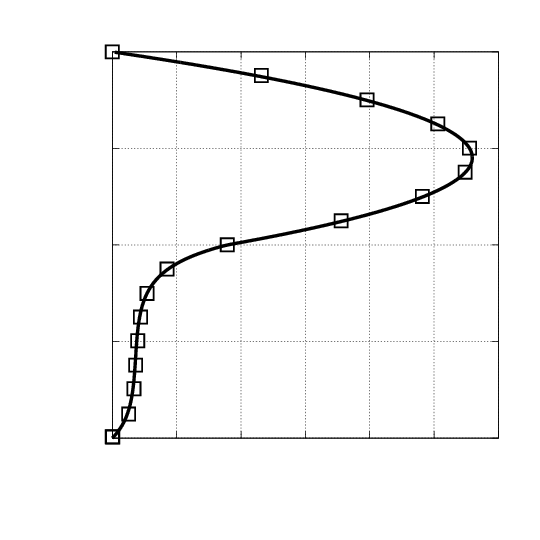}}
        \subcaption{}
        \label{fig:porouschannel-da-02}
    \end{subfigure}
    \hfill  
    \begin{subfigure}{0.48\textwidth}
        \centering
        \scalebox{0.8}{\input{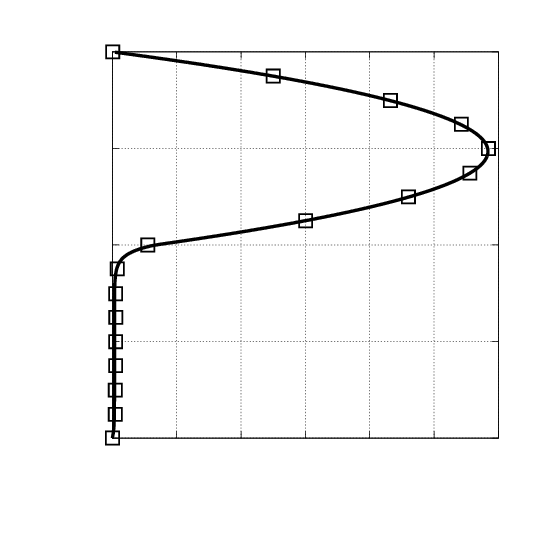}}
        \subcaption{}
        \label{fig:porouschannel-da-03}
    \end{subfigure}   
    \caption{Velocity profile along the $y$ direction: comparison between the simulated  profile (line) and the numerical results of \citet{betchen2006nonequilibrium} (squares) (a) $K=1\times10^{-2}\times H^2$, $U_{in} = 1.17$ (b) $K=1\times10^{-3}\times H^2$, $U_{in} =1.04$}
    \label{fig:porouschannel}
\end{figure*}

\tc{The Beavers and Joseph problem \mbox{\cite{beavers1967boundary}} is a well-known example in fluid mechanics that examines steady plane flow with an interface between a pure fluid and a porous region that lies parallel to the flow direction. A schematic drawing of the computational domain is depicted in \autoref{fig:bej-domain}. The upper region is a free-flow region, whereas the lower region is occupied by a porous medium. Fluid is entering from the left only in the free flow region, but can exit from both regions at the right boundary. The proportions of the domain and the operating conditions were selected following \citet{betchen2006nonequilibrium}. The inlet velocity \mbox{$U_{in}$} is such that, when the profile is fully developed, a condition of \mbox{$Re=1$} is achieved in the free flow region, with a denoted average velocity \mbox{$U$}. \autoref{fig:porouschannel} reports the comparison between the simulated velocity profile and the results obtained by \citet{betchen2006nonequilibrium} for two different permeability values. For both cases, consider a constant porosity \mbox{$\epsilon=0.7$}}
\begin{figure}
    \centering
    \usetikzlibrary{intersections}
\usetikzlibrary{arrows.meta}
\usetikzlibrary{calc}
\definecolor{Lightgreen}{HTML}{87A896}
\definecolor{Darkgreen}{HTML}{004225}
\usetikzlibrary{patterns}

\begin{tikzpicture}[scale=4, font=\small]

  \fill[fill=Lightgreen] (0,0) rectangle +(2,0.25);
  \fill[pattern=big dots, pattern color=white] (0,0) rectangle +(2,0.25);

  \draw[thick] (0,0) rectangle +(2,0.5);

  \draw (0,0) -- +(-0.1,0);
  \draw (0,0.5) -- +(-0.1,0);
  \draw[<->] (-0.05,0) -- (-0.05,0.5);
  \node[left, align=center] at (-0.05,0.25) {\footnotesize inlet\\$2H$};

  \draw (0,0.5) -- +(0,0.1);
  \draw (2,0.5) -- +(0,0.1);
  \draw[<->] (0.,0.55) -- (2,0.55);
  \node[above, align=center] at (1,0.55) {\footnotesize  closed\\$8H$};

  \node[above, right] at (2,0.3333) {\footnotesize outflow};

    \def\numpoints{8}
    \def\A{9}
    \foreach \i in {1, ..., \numpoints} {
      \draw[->] (1.90, \i*0.5/\A) -- +(0.08, 0);
    }
    \foreach \i in {1, ..., 4} {
      \draw[->] (0.05, \i*0.5/\A  + 0.225) -- +(0.08, 0);
      }

\end{tikzpicture}
    \caption{\tc{Computational domain for the Beavers and Joseph Problem}}
    \label{fig:bej-domain}
\end{figure}
\tc {It can be observed that good agreement is found between the two numerical results, highlighting the capability of the proposed framework to be applied to non-reactive test cases. As the permeability decreases, the flow conditions tend to complete blockage of the fluid. Small discrepancies may be related because \citet{betchen2006nonequilibrium} do not explicitly report the actual inlet velocity \mbox{$U_{in}$} used in their simulation; therefore, this value may differ slightly between the two numerical tests. The results obtained are also in line with the numerical work of \citet{costa1992modelling}. In contrast to the previous works, the framework proposed here allows for a moving interface, possibly enabling the study of the time evolution of the system with a receding porous-medium interface.}

\subsection{Mass conservation}\label{sec:mass-conservation}
This section addresses the convergence of mass conservation and the correct calculation of the solid and released gas velocity. 
For this purpose, consider a square computational domain with a spherical homogeneous particle in the bottom left corner, exploiting symmetry conditions on the bottom and left boundaries. 
An initial radius of $R_0$ and isothermal conditions are assumed, so that $\sum_{i}^{NSS}\dot\Omega_{i}$ is constant in time and across the whole solid domain.
Simulations are performed at different grid refinement levels, reported as the number of cells $N$ per length $L$ of the computational domain. 
Additionally, different values of $Z$ are considered to address the effect of this function: pure shrinking regime ($Z = 1$), only porosity increase or fixed interface regime ($Z = 0$) and an in-between situation where both occur at the same time. For the last case, consider a situation where shrinking occurs closer to the surface of the particle, as observed experimentally by \citet{barr2021towards}. Accounting for this can be achieved through a radial $Z$ functional of the type:
\begin{equation}\label{eq:zeta-smooth}
    Z(r) = \frac{1}{1+exp{(32R - 40r)}}  
\end{equation}
where $r$ is the distance between the centre and the point in which $Z$ is being calculated, and $R$ is the current particle radius. This case is referred to as \emph{smooth} because it provides a smooth transition between the two limiting cases: $Z(R)\approx1$ and $Z(0)\approx0$.
\begin{figure*}
    \centering
        \begin{subfigure}{0.48\textwidth}
        \centering
        \scalebox{0.8}{\input{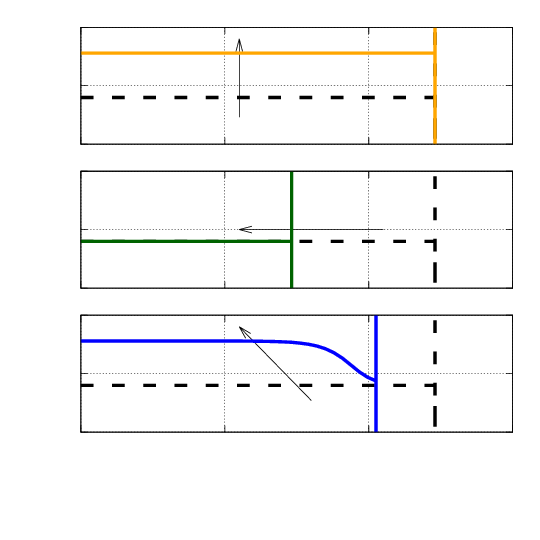}}
        \subcaption[]{}
        \label{fig:massb-radial-porosity}
    \end{subfigure}
    \hfill  
    \begin{subfigure}{0.48\textwidth}
        \centering
        \scalebox{0.8}{\input{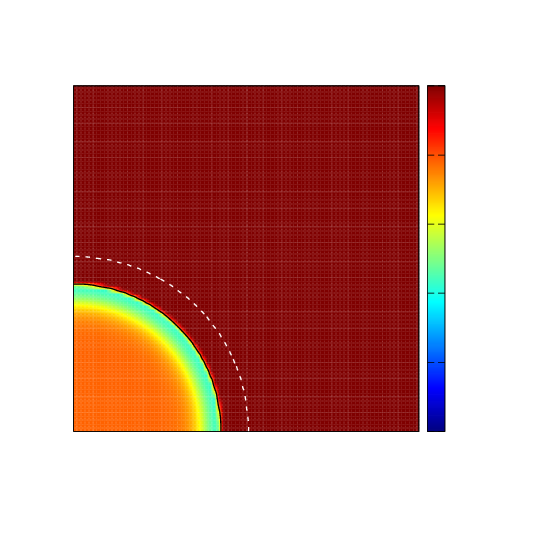}}
        \subcaption[]{}
        \label{fig:massb-porosity-smooth}
    \end{subfigure}    
    \caption{Results for the simulation of the idealised pyrolysis case for the different $Z$ functions. The reported data refer to the simulation performed at the maximum grid refinement case. (a) Radial porosity distribution: the black line represents the initial conditions, while the vertical lines highlight the position of the interfaces. (b) Map of the porosity field at end time for the smooth function case, the dashed line represents the initial position of the interface}
    \label{fig:massb-plots}
\end{figure*}
Simulations were performed for the three different cases: pure shrinking regime, constant volume (or fixed interface) and smooth function.
\autoref{fig:massb-plots} shows the results considering $R_0 = 0.5$ m, $\epsilon_0= 0.4$, $t_{end}=10$ s, $\rho_s=100$ kg m\ts{-3} and sum of formation rates $\sum_{i}^{NSS}\dot\Omega_{i} = - 10$ kg m\ts{-3} s\ts{-1}. 
\autoref{fig:massb-radial-porosity} displays the profiles of the porosity along the radial direction \tc{\mbox{$r$}}.
In the constant volume case ($Z=0$), it is possible to observe that the interface remains fixed while the porosity increases due to chemical reactions. 
On the contrary, in the pure shirking case, the porosity is identical to the initial snapshot, whereas the interface has regressed to preserve mass conservation. In the in-between situation, both phenomena occur simultaneously. 
The interface has regressed, but to a lesser extent, as part of the chemical reaction modified the porosity in the core region rather than solely contributing to particle shrinkage.
Furthermore, a gradient in the porosity field can be observed due to the different values of $Z$ in the two different regions. This result is also emphasised by \autoref{fig:massb-porosity-smooth}, which shows a snapshot at the end time of the simulation for the smooth function case.  

In these simple conditions, the analytical solution of the solid mass loss profile of the particle is also available and follows an equation of the type:
\begin{equation}
    \frac{M_s(t)}{M_s^0} = exp\left(\frac{\sum_{i=1}^{NSS}\dot\Omega_i}{\rho_s}t\right)
\end{equation}
The total produced gas mass profile can be easily recovered, considering that at any time $\frac{M_g(t)}{M_s^0}+\frac{M_s(t)}{M_s^0} = 1$, i.e. the mass moving from the solid phase to the gas phase must be conserved. 
\begin{figure*}
    \centering
        \begin{subfigure}{0.48\textwidth}
        \centering
        \scalebox{0.8}{\input{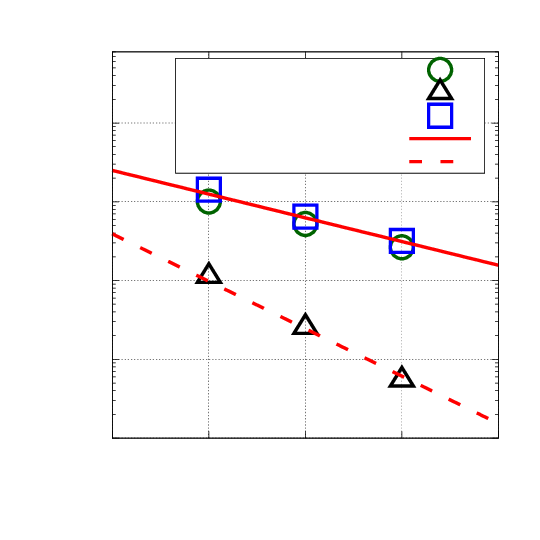}}
        \subcaption{}
        \label{fig:solid-mass-convergence}
    \end{subfigure}
    \hfill  
    \begin{subfigure}{0.48\textwidth}
        \centering
        \scalebox{0.8}{\input{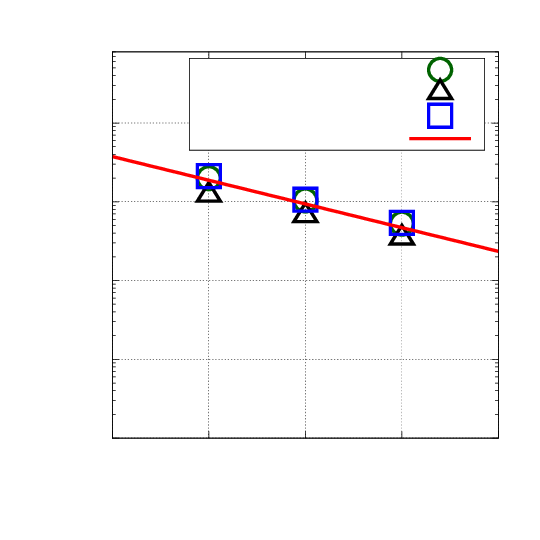}}
        \subcaption{}
        \label{fig:gas-mass-convergence}
    \end{subfigure}    
    \caption{Relative error on the mass conservation at increasing grid refinement. (a) Conservation error in solid mass. (b) Conservation error in produced gas mass. The abscissa represents the maximum number of cells for each dimension of the domain.}
    \label{fig:mass-convergence}
\end{figure*}
\autoref{fig:mass-convergence} shows the error in terms of the absolute difference between the numerical and the analytical solution. 
More in detail, \autoref{fig:solid-mass-convergence} highlights two different behaviours in the order of convergence. In the case of pure porosity variation, second-order convergence is observed, whereas in the pure shrinking and smooth cases, a first-order convergence appears. The explanation for this is related to the fact that, with a fixed interface, the error is associated just with the initialisation of the volume fraction, i.e., the ability of the initialisation method to create a volume fraction that approximates a perfect circle. More accurate initialisation methods, such as the \emph{Vofi} library \cite{bna2016vofi}, were able to reduce the error even further. 
In the other cases, the first-order error appears due to the accuracy of the method of the interface transport method of \citet{WEYMOUTH20102853} when applied to non-divergence-free flows.
Nonetheless, convergence of both the solid and gas mass is achieved, highlighting the strength and correctness of the proposed solution method.

It is clear that the shape and form of $Z$ strongly influence the overall particle shrinkage.
However, this is always constrained between the two limiting cases of pure shrinking and fixed interface.
As mentioned before, the ideal function should depend on the physics of the problem and be in good agreement with the experimental results under different conditions. In \autoref{sec:applications}, a general shape for $Z$ is proposed, and its application is tested against different real cases.

\section{Model validation}\label{sec:applications}
This section highlights the application of the proposed model to non-ideal conditions and its ability to predict yields, shrinking rates, mass degradation and temperature profiles. 

\tc{The proposed mathematical and numerical formulation is independent of the number and shape of biomass particles present. The model can be used to study arbitrary combinations of the latter, as all information is captured in the volume fraction scalar field. However, experimental measurements of interactions among multiple particles under simplified conditions are, to the author's knowledge, unavailable. Therefore, the proposed model is validated on individual particles, for which data are widely available, and will be used in the future to study the effects of such interactions on yields and heating profiles. The possibility of including multiple particles underscores the model's capacity to serve as a sub-grid model finder for larger-scale computational tools.}

In the next simulations, consider the following expression for $Z$:
\begin{equation}\label{eq:zeta-reaction}
    Z=\frac{\sum_{i=i}^{NSS}\dot\Omega_{i}}{\underset{\textit{solid cells}}{max}\left(\sum_{i=1}^{NSS}\dot\Omega_{i}\right)}
\end{equation}
This equation is inspired by the approach proposed by \citet{gentile2017bioSMOKE}, which used a similar formula but in a discontinuous way, changing abruptly between cells with pure shrinking and cells with only porosity variations. Given that the proposed new model can account for both effects at the same time, this form is smoother and leads to fewer numerical oscillations. It should be noted that results are sensitive to the utilised shape of $Z$; however, the strength of the proposed approach lies in the possibility of encompassing both shrinking and porosity variation, while solving the surrounding environment with a moving interface. During model development, it was observed that this function showed an overall particle shrinkage close to $Z=0.5$. However, the form based on the reaction rate was preferred because it leads to a porosity distribution being higher in the core region of the biomass particle. This was found to be in better agreement with the experimental evidence collected by \citet{barr2021towards}, who confirmed the predictions of the model of \citet{gentile2017bioSMOKE}.

\tc{A grid independence study was performed for each of the following cases, so the results presented in the next part are to be considered as not affected by grid resolution. The following results at different grid refinement levels are made available as supplementary materials.}

\tc{ The presented experiment and simulations are conducted under conditions in which the isothermal assumption is invalid, as demonstrated by the internal temperature profiles. This is particularly accentuated due to additional thermal effects arising from pyrolysis reactions and water evaporation, further justifying the need for a spatially resolved multidimensional model.}

Lastly, for all the following test cases, consider:  i) the emissivity of wood for the interface radiation to be $\alpha = 0.9$ \cite{Lopez2013woodemissivity} ii) the solid particle to be initially at ambient temperature $T_p^0 = 298$ K and rapidly inserted into the hot environment so that the dynamics of this step can be neglected.

\subsection{Isotropic sphere}\label{sec:isotropic-sphere}
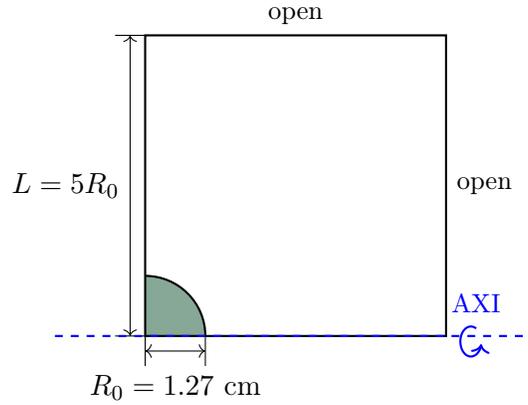
\begin{figure}
    \centering
    \usetikzlibrary{intersections}
\usetikzlibrary{arrows.meta}
\usetikzlibrary{calc}
\definecolor{Lightgreen}{HTML}{87A896}
\definecolor{Darkgreen}{HTML}{004225}

\begin{tikzpicture}[scale=4, font=\small]

  \begin{scope}
    \clip (0,0) rectangle +(1,1);
    \fill[Lightgreen] (0,0) circle (0.2);
    \draw[thick] (0,0) circle (0.2);
  \end{scope}
  \draw[thick] (0,0) rectangle +(1,1);

  \draw (0,0) -- +(-0.1,0);
  \draw (0,1) -- +(-0.1,0);
  \draw[<->] (-0.05,0) -- (-0.05,1);
  \node[left] at (-0.05,0.5) {$L = 5R_0$};

  \draw (0,0) -- (0,-0.1);
  \draw (0.2,0) -- (0.2,-0.1);
  \draw[<->] (0,-0.05) -- (0.2,-0.05);
  \node[below, yshift=-5pt] at (0.1,-0.05) {$R_0 = 1.27$ cm};

  \node[above] at (0.5,1) {\footnotesize open};
  \node[above, right] at (1,0.5) {\footnotesize open};

  \draw[thick, dashed, blue] (-0.3,0) -- (1.3,0);
  \draw[->, blue, thick, shift={(0.75,0)}, xscale=0.7] (0.5,0.025) arc (60:350:0.05);
  \node[above, blue, yshift=5pt] at (1.1,0) {\footnotesize AXI};

\end{tikzpicture}
    \caption{Computational domain for isotropic sphere with imposed surface temperature profile (case \ref{sec:isotropic-sphere})}
    \label{fig:corbetta-domain}
\end{figure}
\begin{figure}
    \centering
    \scalebox{1}{\input{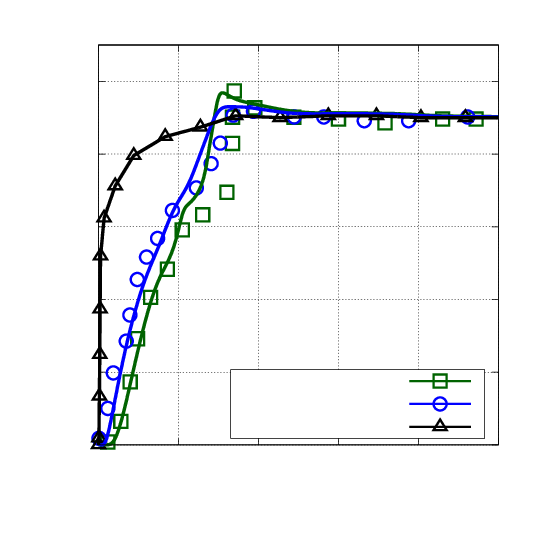}}
    \caption{Temperature profiles inside the particle, experimental (points)\cite{corbetta2014pyrolysis} versus numerical (lines) results}
    \label{fig:corbetta-temperature-profiles}
\end{figure}
\setcounter{table}{0}
\begin{table}[htbp]
\centering
\caption{Properties for test-case \ref{sec:isotropic-sphere} are identical to the ones used by previous numerical works \cite{gentile2017bioSMOKE, corbetta2014pyrolysis}. The reader can find the description of the surrogate species, used as initial composition, in the work by \citet{debiagi2015extractives}.}
\label{tab:corbetta-prop}
\begin{tabularx}{\textwidth}{
 >{\raggedright\arraybackslash}X
 >{\raggedleft\arraybackslash}X
 >{\raggedleft\arraybackslash}X }
     \hline
     Property &  Value & Units \\
     \hline
     Initial radius         & $1.27$   & [cm]\\
     Skeletal density       & $850$                 & [kg m\ts{-3}]\\
     Initial porosity   & $0.4$                 & [-]\\
     Permeability           & $10^{-14}$            & [m\ts{2}]\\
     Thermal conductivity   & $0.1937\times\omega_{\textit{wood}}+0.1405\times\omega_{\textit{char}}+\lambda_g \epsilon$  \cite{corbetta2014pyrolysis} & [W m\ts{-1} K\ts{-1}]\\
    \hline
    Initial composition & & [kg kg\ts{-1}]\\
    \hline
    \footnotesize
    \begin{tabular*}{0.98\textwidth}{@{\extracolsep{\fill}}llllllll@{}}
        CELL & XYHW & LIGO & LIGH & LIGC & TANN & TGL & ASH \\
        0.4807 & 0.2611 & 0.1325 & 0.0957 & 0.0214 & 0.0000 & 0.0000 & 0.0086\\
    \end{tabular*}\\
    \hline
\end{tabularx}
\end{table}

The first test case aims to assess the accuracy of the pseudo-phase solution in terms of heat and mass transfer within the particle and the consequential release of volatiles. Therefore, the focus for this first comparison is only on the particle itself. To achieve this, consider the degradation of a biomass sphere with an imposed temperature profile at the interface. 
\autoref{fig:corbetta-domain} shows a schematic drawing of the computational domain, whereas \autoref{tab:corbetta-prop} summarises the operating conditions and properties used. 
The experimental measurements were performed by \citet{corbetta2014pyrolysis}, where a dried poplar wood sphere initially at room temperature is inserted into a turbulent nitrogen flow at $743$ K. \autoref{fig:corbetta-temperature-profiles} shows the recorded temperature profiles at the core, half radius and surface, which are compared to the results of the numerical solution. 
Considering the high uncertainty and variability involved in biomass pyrolysis, good agreement with the experimental measurements is achieved for both the half radius and core temperature profile. 
As temperatures increase due to heat conduction, chemical reactions initiate throughout the porous structure, causing volatile compounds to be released. The heat absorbed or generated by these reactions subsequently influences the heating profile. Notably, the endothermic deviation of the primary decomposition of the primary wood constituents (cellulose, lignin, hemicellulose, ...) is visible at $\approx 210s$. Additionally, at around $\approx 300s$, it is possible to observe the peak associated with exothermic char formation reactions \cite{milosavljevic1995cellulose}, which makes the internal temperature exceed that of the external environment. By the end, only charred residue is left, and the temperature levels off to match that of the surrounding gas.  
\begin{figure}
    \centering
    \subfloat[] {
    \scalebox{0.7}{\input{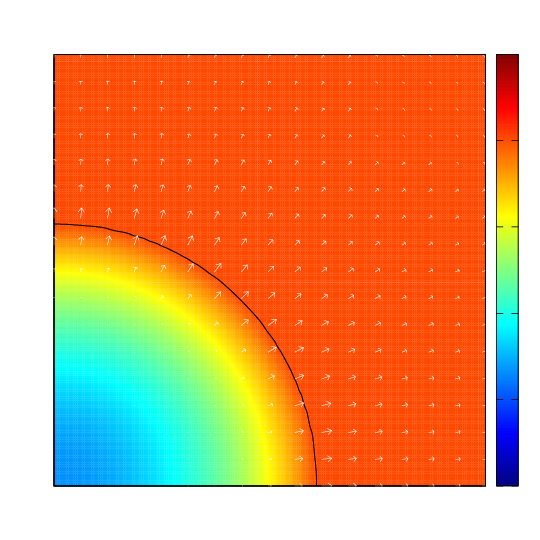}}
    \label{fig:corbetta-100s}
    }\hfill
    \subfloat[] {
    \scalebox{0.7}{\input{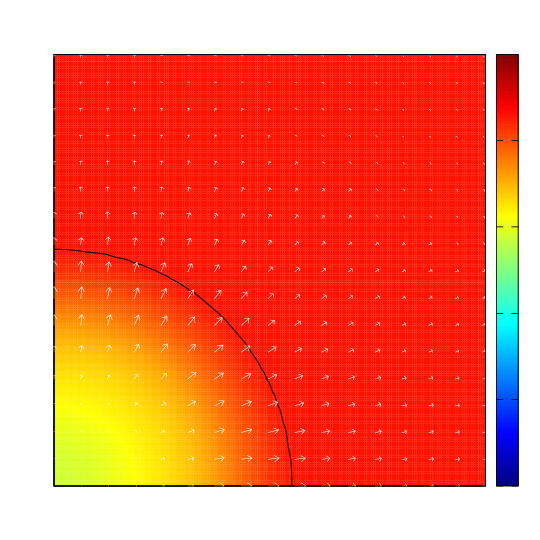}}
    \label{fig:corbetta-200s}
    }\hfill
    \subfloat[]{
    \scalebox{0.7}{\input{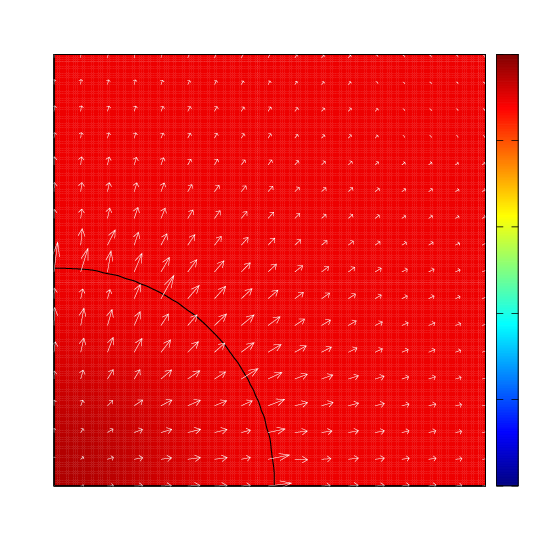}}
    \label{fig:corbetta-300s}
    }\hfill
    \subfloat[]{
    \scalebox{0.7}{\input{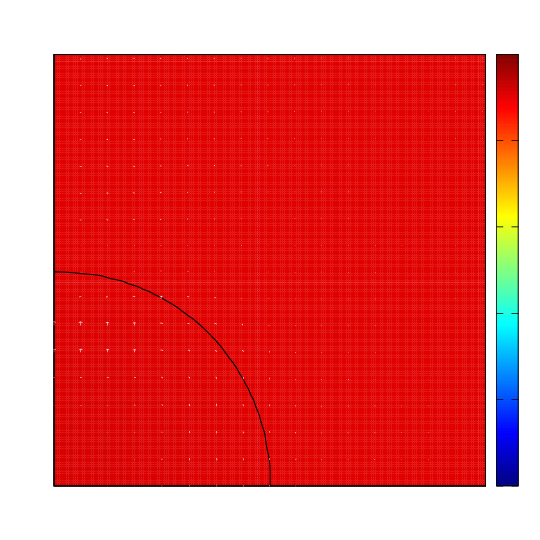}}
    \label{fig:corbetta-600s}
    }\hfill
    \caption{Snapshots of the temperature field for the isotropic sphere case described in \autoref{sec:isotropic-sphere}. The arrows highlight the velocity field. (a) $t=100s$, $|\mathbf{v}_g|_{max}=1.088$ cm s\ts{-1}. (b) $t=200s$, $|\mathbf{v}_g|_{max}=1.138$ cm s\ts{-1}. (c) $t=300s$, $|\mathbf{v}_g|_{max}=3.406$ cm s\ts{-1} (d) $t=600s$, $|\mathbf{v}_g|_{max}=0.365$ cm s\ts{-1}.}
    \label{fig:corbetta-snapshots}
\end{figure}
\autoref{fig:corbetta-snapshots} shows some zoomed snapshots of the temperature field at different times, together with arrows showcasing the gas flow field. Volatile release starts from the outer region and progresses through the particle, following the temperature evolution (\autoref{fig:corbetta-100s} and \ref{fig:corbetta-200s}). \autoref{fig:corbetta-300s} also highlights the continuity of the gas velocity across the interface. Little numerical oscillations are observed in the interfacial cells due to the sharp variation of the Darcy drag term across it. Although subtle, the core temperature overshoot is also visible in \autoref{fig:corbetta-300s}.

\begin{figure}
    \centerline{\scalebox{0.7}{\input{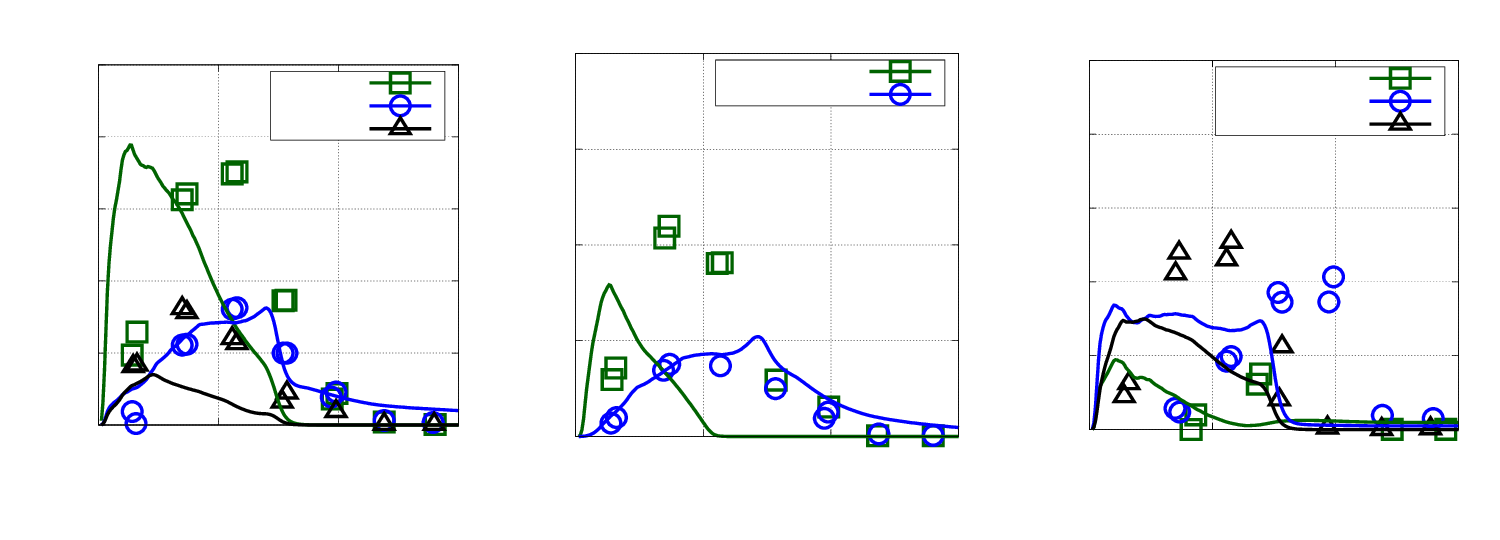}}}
    \caption{Released volatiles from wood sphere pyrolysis \ref{sec:isotropic-sphere}. Comparison between numerical (lines) and experimental (points)\cite{corbetta2014pyrolysis} results.}
    \label{fig:corbetta-species}
\end{figure}

Experimental measurements of the released oxygenated compounds were also available and therefore compared with the simulated volatiles released crossing the interface at a given time. As displayed in \autoref{fig:corbetta-species}, good agreement is found both in terms of release rate and order of magnitude for each species. During the degradation, different peaks appear as the production of one volatile is correlated to the degradation of multiple starting compounds. The degradation of the latter does not occur at the same time/temperature, explaining the different peaks observed \cite{Ranzi2017-vu}. Overall, some but reasonable discrepancies can be observed, especially for CH\textsubscript{3}COOH and HCOOH. It should be stated that these results are sensitive to the kinetic mechanism utilised. Nonetheless, the generality of the model allows different schemes in CHEMKIN \cite{kee1996chemkin} format to be utilised, as mentioned in \autoref{sec:numerical-strategy}.

\subsection{Wooden sphere in flow}\label{sec:huang-sphere}
\begin{figure*}
    \centering
    \centerline{\begin{subfigure}{0.48\textwidth}
        \centering
        \scalebox{0.8}{\input{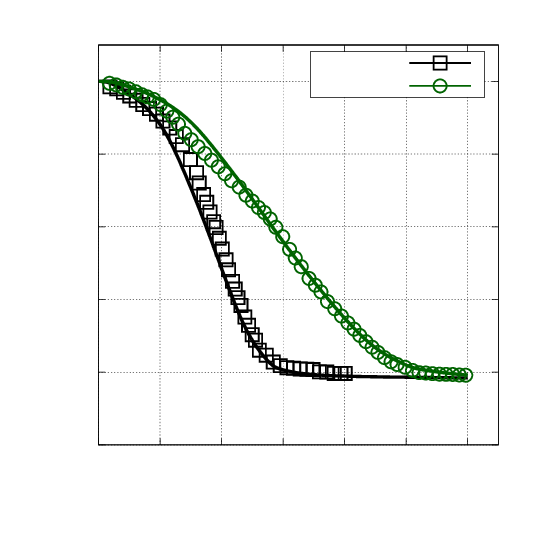}}
        \subcaption{}
        \label{fig:huang-mass}
    \end{subfigure}
    \hfill  
    \begin{subfigure}{0.48\textwidth}
        \centering
        \scalebox{0.8}{\input{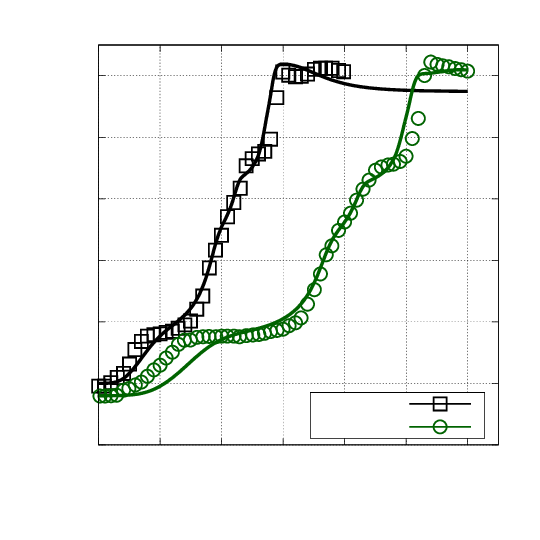}}
        \subcaption{}
        \label{fig:huang-t-core}
    \end{subfigure}}
    \caption{Results at reactor working temperature equal to $773$ K for the two particle diameters investigated. (a) Mass loss profile. (b) Core temperature profile. Comparison between experimental (points)\cite{huang2014modeling} and numerical (lines).}
    \label{fig:huang-mass-and-t-core}
\end{figure*}

This test case removes the constraint of the imposed interface temperature profile, resolving the coupled heat and mass transfer between the particle and its external environment. Therefore, sub-grid-scale correlations are no longer needed, and exchange terms are calculated through direct solution of the interface boundary layers. The experimental apparatus and data are proposed by \citet{huang2014modeling}. The authors analysed the evolution of Schima superba wood spheres inserted into a laminar hot nitrogen flow at different reactor temperatures and for two sphere diameters ($2$ and $3$ cm). 
\begin{figure}
    \centering
    \usetikzlibrary{intersections}
\usetikzlibrary{arrows.meta}
\usetikzlibrary{calc}
\definecolor{Lightgreen}{HTML}{87A896}
\definecolor{Darkgreen}{HTML}{004225}

\begin{tikzpicture}[scale=4, font=\small]

  \begin{scope}
    \clip (0,0) rectangle +(2, 0.6666);
    \fill[Lightgreen] (1,0) circle (0.3333);
    \draw[thick] (1,0) circle (0.3333);
  \end{scope}
  \draw[thick] (0,0) rectangle +(2,0.6666);

  \draw (0,0) -- +(-0.1,0);
  \draw (0,0.6666) -- +(-0.1,0);
  \draw[<->] (-0.05,0) -- (-0.05,0.6666);
  \node[left, align=center] at (-0.05,0.3333) {\footnotesize inlet\\$L=4$ cm};

  \draw (0.6667,0) -- +(0,-0.1);
  \draw (1.3333,0) -- +(0,-0.1);
  \draw[<->] (0.6667,-0.05) -- (1.3333,-0.05);
  \node[below, yshift=-5pt] at (1,-0.05) {$D_0$};

  \draw (0,0.6666) -- +(0,0.1);
  \draw (2,0.666) -- +(0,0.1);
  \draw[<->] (0.,0.7166) -- (2,0.7166);
  \node[above, align=center] at (1,0.7166) {\footnotesize  hot wall, closed\\$3L$};

  \node[above, right] at (2,0.3333) {\footnotesize outflow};

  \draw[thick, dashed, blue] (-0.3,0) -- (2.3,0);
  \draw[->, blue, thick, shift={(0.75,0)}, xscale=0.7] (1.95, 0.025) arc (60:350:0.05);
  \node[above, blue, yshift=5pt] at (2.1,0) {\footnotesize AXI};

    \def\numpoints{6}
    \def\A{7}
    \foreach \i in {1, ..., \numpoints} {
      \draw[->] (0.02, \i*0.6666/\A) -- +(0.08, 0);
      \draw[->] (1.90, \i*0.6666/\A) -- +(0.08, 0);
    }

\end{tikzpicture}
    \caption{Computational domain for the isotropic sphere in flow (case \ref{sec:huang-sphere}). The drawing proportions refer to $D_0 = 2$ cm}
    \label{fig:hunag-domain}
\end{figure}
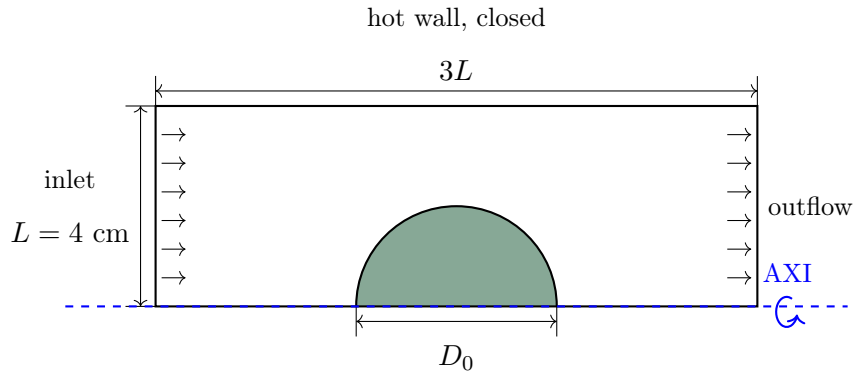
\autoref{fig:hunag-domain} shows a schematic drawing of the computational domain. 
\autoref{tab:hunag-prop} presents the physical properties employed for this test case, which largely correspond to those documented by the authors alongside their experimental measurements in the same study. 
Note that the authors did not report/perform an ultimate analysis on the wood sample utilised. Given that this information is essential to estimate the initial surrogate wood composition, average data from similar wood samples were used.
The particles were not oven-dried but contained $10\%$ of moisture, which significantly alters the heating profile.
\begin{table}[htbp]
\centering
\caption{Properties for test-case \ref{sec:huang-sphere}. The reader can find the description of the surrogate species, used as initial composition, in the work by \citet{debiagi2015extractives}.}
\label{tab:hunag-prop}
\begin{tabularx}{\textwidth}{
>{\raggedright\arraybackslash}X
>{\raggedleft\arraybackslash}X
>{\raggedleft\arraybackslash}X }
    \hline
    Property &  Value & Units \\
    \hline
    Initial diameter       & $2.0$ or $3.0$   & [cm]\\
    Skeletal density       & $920$                 & [kg m\ts{-3}]\\
    Initial gas porosity   & $0.39$                 & [-]\\
    Permeability           & $10^{-14}$            & [m\ts{2}]\\
    Thermal conductivity   & $0.210\times\omega_{\textit{wood}}+0.071\times\omega_{\textit{char}}+\lambda_g \epsilon_g$ \cite{DIBLASI19961121}& [W m\ts{-1} K\ts{-1}]\\
   \hline
   Initial composition & & [kg kg\ts{-1}]\\
   \hline
    \footnotesize
    \begin{tabular*}{0.98\textwidth}{@{\extracolsep{\fill}}lllllllll@{}}
        CELL & XYHW & LIGO & LIGH & LIGC & TANN & TGL & ASH & MOIST \\
        0.4229 & 0.1830 & 0.1759 & 0.0364 & 0.0081 & 0.0362 & 0.0245 & 0.0130 & 0.1000\\
    \end{tabular*}\\
   \hline
\end{tabularx}
\end{table}
Simulations were carried out at four different reactor operating temperatures: $673$ K, $773$ K, $873$ K and $973$ K. We focus initially on the cases at $773$ K, as these are those for which the core temperature and mass profiles are reported; the other working temperatures showed a very similar numerical behaviour, and the same is expected for their experimental measurements. \autoref{fig:huang-mass-and-t-core} shows the time evolution of the mass loss profile and the core temperature for the two different analysed initial particle diameters. It can be seen that excellent agreement is achieved between the experimental and numerical results. Looking at the temperature profile, we can notice that the primary degradation endothermic peak and the secondary exothermic peak are present, similarly to the behaviour observed for the previous test case (\ref{sec:isotropic-sphere}).
By contrast, since the wood samples were not pre-dried, a new feature appears around $\approx373$ K, namely a pronounced deviation from the expected temperature rise induced by external heating.
This is caused by the evaporation of moisture, which absorbs latent heat and thereby tends to reduce particle temperature.
It is worth mentioning that its effect is usually considered using an irreversible reaction for the desorption of water \cite{anca2017online, afessa2025pyrolysis}. Using this approach, the internal gas reaches oversaturated, unfeasible conditions, especially within the pores of the biomass. To avoid this, the reverse reaction for the resorption water must also be considered, reducing the evaporation rate and the quenching contribution at temperatures below the boiling point.
Without considering this reversible reaction, the evaporation rate inside the thick particle is too fast, and therefore, the temperature plateau at $\approx373$ K is not well captured.  
This reaction is proposed in this work as an extension to the kinetic mechanism of \citet{debiagi2024cellulose}, and it is provided in the supplementary material to this work as mentioned previously.

\begin{figure*}
    \centering
    \centerline{\begin{subfigure}{0.48\textwidth}
        \centering
        \scalebox{0.8}{\input{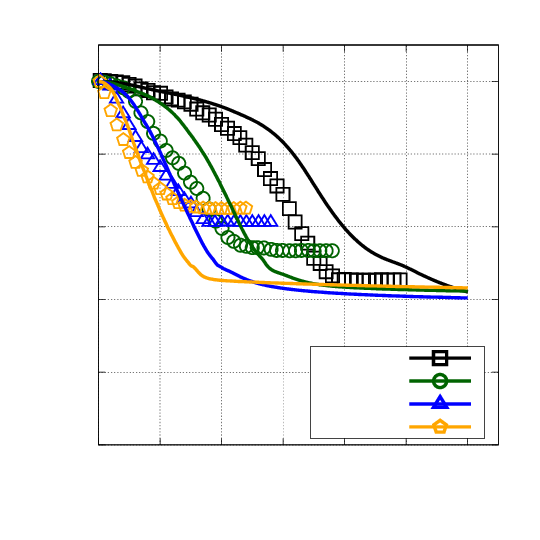}}
        \subcaption{}
        \label{fig:huang-shrink-20}
    \end{subfigure}
    \hfill  
    \begin{subfigure}{0.48\textwidth}
        \centering
        \scalebox{0.8}{\input{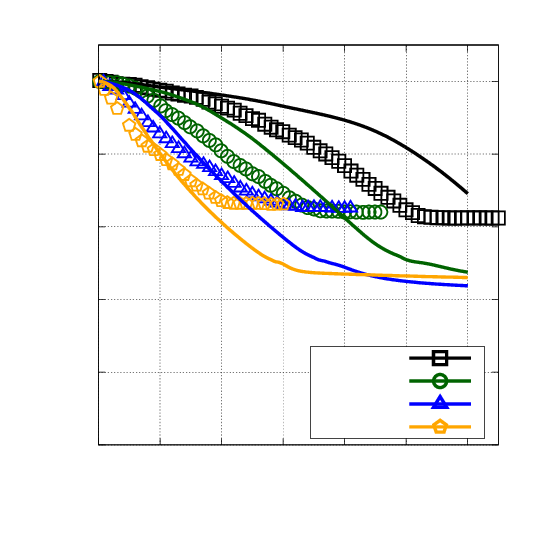}}
        \subcaption{}
        \label{fig:huang-shrink-30}
    \end{subfigure}}
    \caption{Shrinking profiles at different reactor temperatures and different initial diameters: (a) $D_0=2$ cm. (b) $D_0=3$ cm. Comparison between experimental (points)\cite{huang2014modeling} and numerical (lines).}
    \label{fig:huang-shrink}
\end{figure*}

Experimental measurements of particle diameter evolution were available at all operating temperatures and particle diameters.
\autoref{fig:huang-shrink} shows the comparison between experimental and numerical results.
At the smaller diameter (\autoref{fig:huang-shrink-20}), it can be observed that the model under-predicts the dependence of the final shrinking factor on the temperature. However, this behaviour seems to disappear for the larger particle, as seen in \autoref{fig:huang-shrink-30}, showing a trend more in line with the model's prediction. It is not clear why this tendency disappears for the bigger particles; further experimental measurements are necessary for a better understanding of this phenomenon.
Nonetheless, the model correctly captures the experimental trend, predicting a sharp initial reduction in shape as the main components start to degrade. At lower temperatures, such as the case at $673$ K for the $2$ cm particle, the degradation is less pronounced in the first instants but increases once the particle reaches sufficiently high temperatures for all the main constituents to start degrading. As the particle reaches complete conversion, the shrinking stops, reaching a final diameter. The model overestimates the duration required for the shrinking factor to reach the final diameter and the corresponding value. However, the computed average error for the steady state value remains around $\approx10\%$, which is reasonable given the simplified description of the shrinking problem and the high uncertainties and variability involved, including the correct biomass composition and possible effects of its anisotropy.

\subsection{Anisotropic cylinder in forced convection}\label{sec:cylinder}
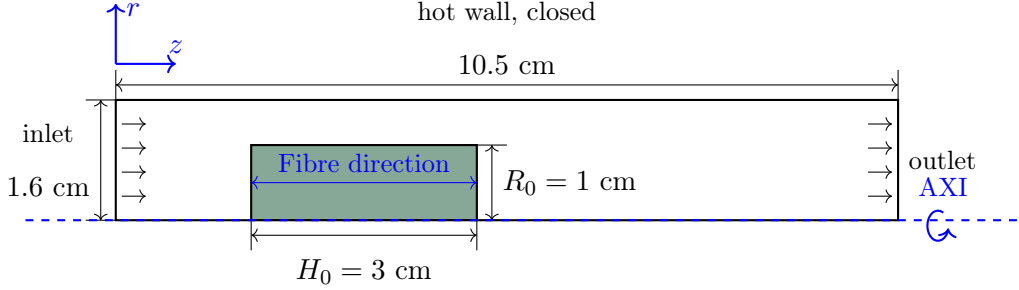
\begin{figure}
    \centering
    \usetikzlibrary{intersections}
\usetikzlibrary{arrows.meta}
\usetikzlibrary{calc}
\definecolor{Lightgreen}{HTML}{87A896}
\definecolor{Darkgreen}{HTML}{004225}

\begin{tikzpicture}[scale=4, font=\small]

  \begin{scope}
    \clip (0,0) rectangle +(2.6,0.4);
    \fill[Lightgreen] (0.45,0) rectangle (1.2, 0.25);
    \draw[thick] (0.45,0) rectangle (1.2, 0.25);
  \end{scope}
  \draw[thick] (0,0) rectangle +(2.6, 0.4);

  \draw (0,0) -- +(-0.1,0);
  \draw (0,0.4) -- +(-0.1,0);
  \draw[<->] (-0.05,0) -- (-0.05,0.4);
  \node[left, align=center] at (-0.05,0.2) {\footnotesize inlet\\$1.6$ cm};

  \draw (1.2,0) -- +(0.1,0.);
  \draw (1.2,0.25) -- +(0.1,0);
  \draw[<->] (1.25,0) -- (1.25,0.25);\
  \node[right] at (1.25, 0.125) {$R_0=1$ cm};

  \draw (0,0.4) -- +(0,0.1);
  \draw (2.6,0.4) -- +(0,0.1);
  \draw[<->] (0,0.45) -- (2.6,0.45);
  \node[above, align=center] at (1.3,0.45) {\footnotesize hot wall, closed\\$10.5$ cm};

  \draw (0.45,0) -- (0.45,-0.1);
  \draw (1.2,0) -- (1.2,-0.1);
  \draw[<->] (0.45,-0.05) -- (1.2,-0.05);
  \node[below, yshift=-5pt] at (0.825,-0.05) {$H_0 = 3$ cm};

  \node[right] at (2.6,0.2) {\footnotesize outlet};

  \draw[thick, dashed, blue] (-0.3,0) -- (3,0);
  \draw[->, blue, thick, shift={(2.4,0)}, xscale=0.7] (0.5,0.025) arc (60:350:0.05);
  \node[above, blue, yshift=5pt] at (2.75,0) {\footnotesize AXI};

  \draw[thick, blue, ->] (0,0.52) -- +(0.2,0);
  \draw[thick, blue, ->] (0,0.52) -- +(0,0.2);
  \node[above, blue] at (0.2, 0.52) {$z$};
  \node[right, blue] at (0, 0.7) {$r$};

  \draw[->] (0.02, 0.08) -- +(0.08, 0);
  \draw[->] (0.02, 0.16) -- +(0.08, 0);
  \draw[->] (0.02, 0.24) -- +(0.08, 0);
  \draw[->] (0.02, 0.32) -- +(0.08, 0);

  \draw[->] (2.5, 0.08) -- +(0.08, 0);
  \draw[->] (2.5, 0.16) -- +(0.08, 0);
  \draw[->] (2.5, 0.24) -- +(0.08, 0);
  \draw[->] (2.5, 0.32) -- +(0.08, 0);

  \draw[<->, blue] (0.45, 0.125) -- (1.2, 0.125);
  \node[above, blue] at (0.825, 0.125) {\footnotesize Fibre direction};

\end{tikzpicture}
    \caption{Computational domain for the anisotropic cylinder (case \ref{sec:cylinder})}
    \label{fig:cylinder-domain}
\end{figure}
 This test case showcases the ability of the model to work with more complex particle geometries (e.g. when dealing with whole wood chunks) and with materials which show preferential heat transfer in one direction. To achieve this, we consider the pyrolysis of a beech wood cylinder exposed to a hot stream. The sample is placed in a small reactor so that there is a relevant presence of a radiating hot wall at the top boundary. Nitrogen is introduced from the left side of the domain already at the operating temperature. Axial symmetry is exploited to reduce computational times.
\autoref{fig:cylinder-domain} shows a schematic drawing of the computational domain. 
Gauthier et al.~\citep{GAUTHIER2013521,gauthier2013syntesis} performed the experimental measurements at a reactor/ambient temperature of $723$ K. 
For this case, the anisotropic properties of wood are considered. The thermal conductivity is higher along the fibres of the wood sample, which are aligned with the $z$ coordinate. The values in the tangential and radial direction are estimated according to the approach proposed by \citet{kollmann2012principles}:
\begin{equation}
    \boldsymbol{\lambda}_{m} = \theta\lambda_{||} + (1-\theta)\lambda_{\perp}
\end{equation}
\begin{equation}
    \lambda_{||} = (1-\epsilon_g)\lambda_{s,||} + \epsilon_g\lambda_g
\end{equation}
\begin{equation}
    \lambda_{\perp} = \frac{1}{\frac{1-\epsilon_g}{\lambda_{s,\perp}}+\frac{\epsilon_g}{\lambda_g}} 
\end{equation}
Where: $\boldsymbol{\lambda}_{m}$ is the effective thermal conductivity, $\lambda_{s,||} = 0.766$ W m\ts{-1} K\ts{-1} is the solid thermal conductivity of the fibres along the grain \cite{maku1954studies}, $\lambda_{s,\perp} = 0.430$ W m\ts{-1} K\ts{-1} is the solid thermal conductivity perpendicular to the grain \cite{siau2012transport}, $\lambda_g$ is the gas thermal conductivity (computed by OpenSMOKE++ \cite{CUOCI2015237}) and $\theta$ is a weighting bridge-factor eqaual to $1$ in the longitudianl direction and $0.58$ along the perpendicular direction.
Calculation showed that the expected conductivity in the fibres' direction is roughly twice that in the perpendicular direction. 
For further reproducibility of this test case,\autoref{tab:cylinder-prop} reports a summary of the physical properties utilised. 
Previous numerical works have already simulated this test case \cite{gentile2017bioSMOKE, PATO2014}. However, since these models focused only on the wood particle and not the external environment, they relied on correlations to close the system of equations at the interface. In this work, this assumption is relaxed, solving the surrounding boundary layer to determine the exchanged fluxes between the biomass particle and the external gaseous phase.

The cylinder-shaped biomass is not initialised as a perfect cylinder, with sharp edges, but rather as a superquadric shape \cite{Jaklic2010superquadrics}. This family of geometric shapes is described by the following equation (in 2D): $\left\vert\frac{x}{A}\right\vert^n+\left\vert\frac{y}{B}\right\vert^m=1$. Where $A$ and $B$ are used to regulate the dimension in the $x$ and $y$ direction, while $n$ and $m$ regulate the roundness of the corners. Increasing the exponents leads to sharper corners, eventually tending to the perfect cylinder case.  For this simulation, consider $A=H_0/2$, $B=R_0$ and $n=m=20$. This approach avoids sharp edges in the initialisation of the VOF fraction, which can lead to inaccuracies in the calculation of the interface normals for the Piecewise Linear Interface Calculation (PLIC) reconstruction \cite{UBBINK1999sharpinterfaces}. 
\begin{table}[htbp]
\centering
\caption{Properties for test-case \ref{sec:cylinder} are identical to the ones used by previous numerical works \cite{gentile2017bioSMOKE, Pyromech}. The reader can find the description of the surrogate species, used as initial composition, in the work by \citet{debiagi2015extractives}.}
\label{tab:cylinder-prop}
\begin{tabularx}{\textwidth}{
>{\raggedright\arraybackslash}X
>{\raggedleft\arraybackslash}X
>{\raggedleft\arraybackslash}X }
    \hline
    Property &  Value & Units \\
    \hline
    Initial dimensions     & $z:3.0\quad r:1.0$   & [cm]\\
    Skeletal density       & $1200$                 & [kg m\ts{-3}]\\
    Initial gas porosity   & $0.4$                 & [-]\\
    Permeability           & $z:10^{-14}\quad r:10^{-12}$            & [m\ts{2}]\\
    Thermal conductivity   & \citet{kollmann2012principles} & [W m\ts{-1} K\ts{-1}]\\
    \hline
   Initial composition & & [kg kg\ts{-1}]\\
   \hline
    \footnotesize
    \begin{tabular*}{0.98\textwidth}{@{\extracolsep{\fill}}llllllll@{}}
        CELL & XYHW & LIGO & LIGH & LIGC & TANN & TGL & ASH \\
        0.4169 & 0.3147 & 0.1039 & 0.0595 & 0.0005 & 0.0616 & 0.0349 & 0.0080\\
    \end{tabular*}\\
   \hline
\end{tabularx}
\end{table}

\begin{figure*}
    \centering
    \subfloat[] {
    \centerline{\scalebox{1}{\input{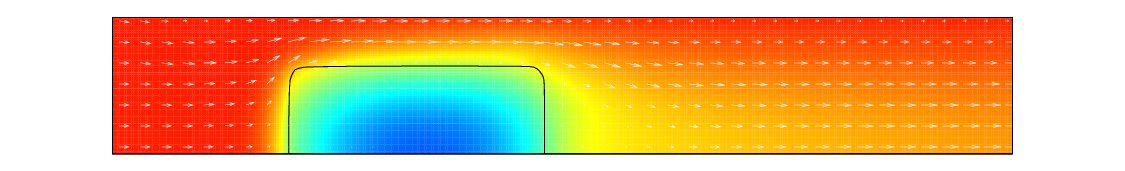}}}
    \label{fig:cylinder-100s}
    }\hfill
    \subfloat[] {
    \centerline{\scalebox{1}{\input{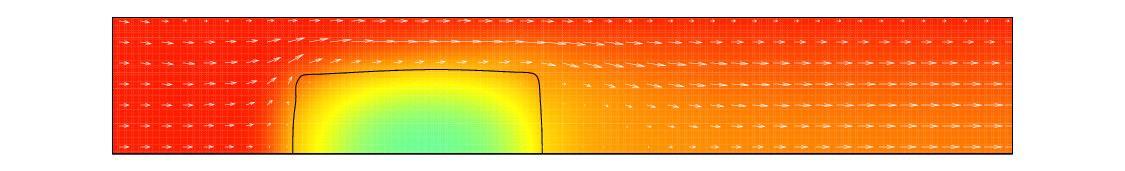}}}
    \label{fig:cylinder-200s}
    }\hfill
    \subfloat[] {
    \centerline{\scalebox{1}{\input{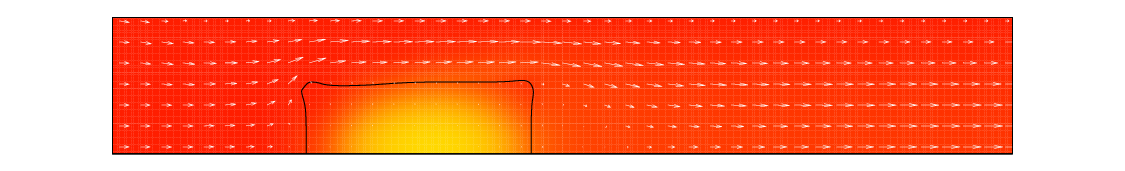}}}
    \label{fig:cylinder-300s}
    }\hfill
    \subfloat[] {
    \centerline{\scalebox{1}{\input{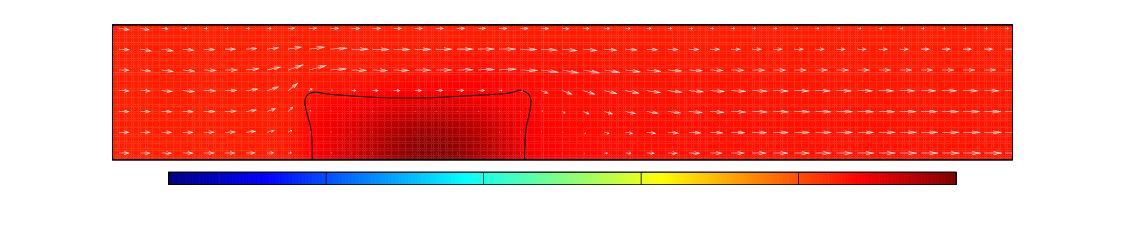}}}
    \label{fig:cylinder-400s}
    }\hfill
    \caption{Snapshots of the temperature field for the anisotropic cylinder case described in \autoref{sec:cylinder}. The arrows highlight the velocity field. (a) $t=100$ s. (b) $t=200$ s. (c) $t=300$ s. (d) $t=400$ s.}
    \label{fig:cylinder-snapshots}
\end{figure*}
\autoref{fig:cylinder-snapshots} shows some temperature field snapshots at different times of the simulation. It can be seen that heating starts from the corners, which also begin to shrink sooner than the central part of the edges.
The particle heating is also slightly de-centred due to the effect of the hot nitrogen flow entering from left to right. The temperature overshoot is also visible at the particle core in \autoref{fig:cylinder-400s}. Ultimately, the particle exhibits a \emph{bone-like} shape, which was also observed in some experiment \cite{Gronliphdthesis} and previous numerical works \cite{gentile2017bioSMOKE}. 
This behaviour is correlated to the earlier degradation of the corners with respect to the middle sections, mainly related to the higher heat flux received in these regions. 

\begin{figure*}
    \centering
    \centerline{\begin{subfigure}{0.48\textwidth}
        \centering
        \scalebox{0.8}{\input{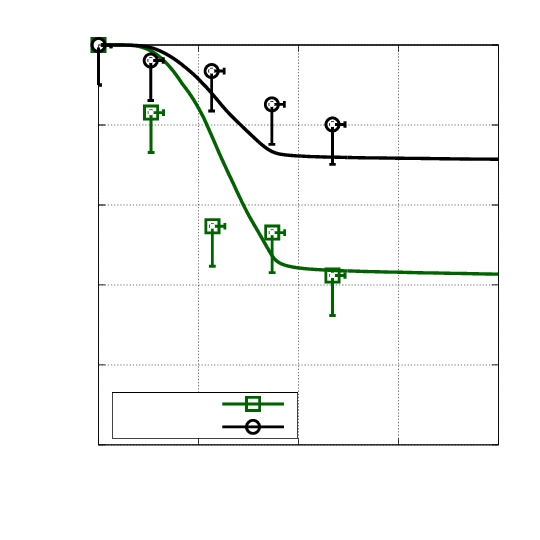}}
        \subcaption{}
        \label{fig:cylider-shrink}
    \end{subfigure}
    \hfill  
    \begin{subfigure}{0.48\textwidth}
        \centering
        \scalebox{0.8}{\input{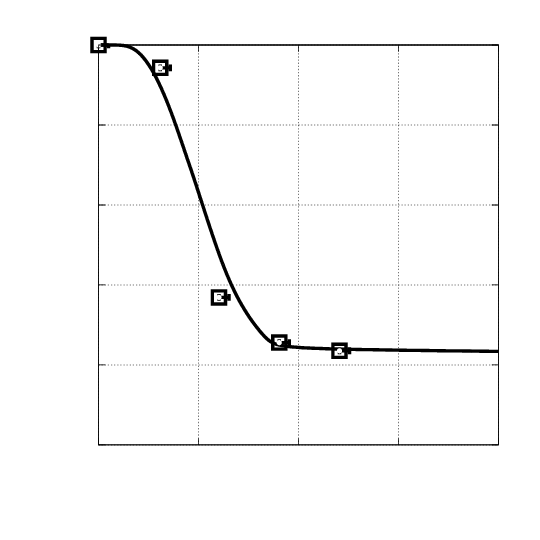}}
        \subcaption{}
        \label{fig:cylidner-mass}
    \end{subfigure}}
    \caption{(a) Shrinking factor along the main directions. (b) Solid mass time evolution. Comparison between experimental (points) \cite{gauthier2013syntesis} and numerical (lines) results for the wood cylinder pellet.}
    \label{fig:cylinder-evolution}
\end{figure*}

\autoref{fig:cylinder-evolution} shows quantitatively the evolution of the particle shape and mass as pyrolysis progresses. \autoref{fig:cylider-shrink} shows the time evolution of the shrinking factor, defined as the current over the initial dimension, along the radial $r$ and axial $z$ directions. As for the previous test cases, a delay can be noticed before the degradations start, due to the rise of temperature before reaching conditions for the pyrolysis reactions to be significant. Once the onset point is reached, the mass profile and particle dimensions sharply decrease.
It can be seen that more shrinking is exhibited along the radial direction. This is mainly due to two factors: the presence of the hot wall at the top boundary, which increases heat transfer along the radial direction, and the anisotropic nature of the thermal conductivity. As this value is lower along the radial direction, heat transfer is hindered along this coordinate, leading to faster heating rates in the topmost cells.
The model slightly underestimates the initial shrinking in both directions, probably due to some underestimation of the heat exchange in the first instants, related also to the sample insertion procedure. Nonetheless, the steady state values are accurately captured, considering the high variability and uncertainties in the properties of these systems. 
\autoref{fig:cylidner-mass} shows that the model is in good agreement with the experimental measurements for the mass loss profile. Additionally, the final value of the remaining mass, a key indicator of the biochar yield of the process, is captured well.

To further validate our model, we compare the calculated interface heat exchange coefficients from our temperature field solution, in the first instants of the simulation, with those estimated by \citet{GAUTHIER2013521} from the experimental temperature profiles. For pure convection (i.e. excluding radiation), our simulation yields $h_{cond} = 12$ W m\ts{-2} K\ts{-1}, compared to $15$ W m\ts{-2} K\ts{-1} from experiments. When radiation is included, the heat exchange coefficient increases to 46 W m\ts{-2} K\ts{-1} in our numerical simulation, compared to 42 W m\ts{-2} K\ts{-1} in the experimental study. The discrepancy appears because \citet{GAUTHIER2013521} uses a linearising function to estimate the radiative contribution. This comparison demonstrates good agreement between our model and experiments, while also confirming the significant role of radiation in the particle's heating behaviour and the model's capacity to serve as an exploration tool for sub-grid correlations.

\section{Conclusion}\label{sec:conclusions}
This work proposes a novel methodology for describing the evolution of biomass particle pyrolysis. The framework relaxes the constraints of a fixed particle interface and the need for heat and mass transfer coefficient correlations, which are often employed as tunable parameters to match experimental measurements. Instead, these effects are computed directly by solving the surrounding flow field, while accounting for an arbitrarily moving interface. To achieve this, a Volume-Of-Fluid model was modified to describe the evolution of a porous phase, incorporating pyrolysis chemical kinetics, Darcy–Forchheimer contributions, and anisotropic transport properties.
The model \tc{\sout{demonstrated} showed} first-order convergence in \tc{particle} mass conservation \tc{, due to the non-divergence-free velocity and the VOF interface transport. Rise to second order is observed for stationary interfaces, where volume fraction initialisation controls the convergence rate.} Comparisons with experimental data showed that the model captures temperature and mass loss profiles \tc{\sout{accurately} with decent accuracy}, while the release of volatiles follows the \tc{\sout{correct}} experimental trends.
The proposed approach for distributing porosity changes and volume variation yields reasonable trends, but also highlights a lack of fundamental understanding regarding the evolution of the internal particle structure. The model can be used to predict yields and characteristic degradation times, aiding the development of pyrolysis processes even under complex fluid dynamic conditions.
\tc{While providing insights into the degradation behaviour of biomass, the model can also serve as a tool for multi-scale modelling of biomass conversion processes. In particular, this model enables detailed studies of single particles of arbitrary geometry under controlled conditions. Additionally, the proposed formulation enables the study of interactions among multiple particles. From these studies, drag, heat, and mass transfer coefficients characterising the systems can be obtained, as well as information on the interplay between porosity and volume changes. This information can be transformed into sub-grid-scale correlations, which are necessary for laboratory- or industrial-scale reactor simulations. Euler-Lagrange \cite{KONG2022131847} or Euler-Euler \cite{tsekos2022two} models, which do not resolve the interface between the particle and the gas phase, usually employ these correlations. This hierarchical strategy for simulating the biomass gasification process will ultimately enable optimisation of gasification and pyrolysis processes with respect to conversion, yields, operating conditions, efficiency, and emissions.}
The model was developed within the open-source code Basilisk, and released in open-source form, in order to enhance the reproducibility of the results, and allowing improvements from the community.

Future work will focus on exploring the effect of gas-phase reactions to test the model's behaviour during combustion and gasification processes. This introduces new challenges, as the number of species required to analyse these systems increases drastically, necessitating chemistry acceleration techniques. These enhancements will enable the prediction of pollutant formation, such as soot and NOx, thereby assessing the environmental impact of these processes. Overall, the proposed model highlights new perspectives in the description of biomass pyrolysis processes, providing insights into the behaviour of such a complex problem and allowing the development of improved engineering correlations.
\section{CRediT authorship contribution statement}
\textbf{Riccardo Caraccio}: Writing – original draft, Software, Validation, Investigation, Formal analysis, Methodology. \textbf{Edoardo Cipriano}: Writing – original draft, Software, Methodology. \textbf{Alessio Frassoldati}: Writing – review \& editing, Conceptualization. \textbf{Tiziano Faravelli}: Writing – review \& editing, Conceptualization,  Funding acquisition.
\section{Acknowledgments}
Funded by the European Union - Next Generation EU, Mission 4 Component 1 CUP D43C23002570001 and by the European Union’s Horizon Europe research and innovation programme under the HORIZON-CL5-2022-D3-02-PYSOLO grant agreement No 101118270

\emph{Disclaimer}: Views and opinions expressed are however those of the author(s) only and do not necessarily reflect those of the European Union or European Climate, Infrastructure and Environment Executive Agency. Neither the European Union nor the granting authority can be held responsible for them.

\bibliographystyle{elsarticle-num-names} 
\bibliography{cas-refs}

\end{document}